\documentclass[oldversion]{aa}  
\usepackage{authblk}
\usepackage{graphicx,amsmath}
\usepackage{amssymb}
\usepackage{color}
\usepackage{natbib}		%for A&A : reimplements the LaTeX \cite command 
\usepackage{url}
\usepackage{multirow}
\usepackage{threeparttable}   %follow the A&A style
\usepackage{soul,mathabx}
\bibpunct{(}{)}{;}{a}{}{,} % to f
\usepackage{lscape}

\begin{document}

   \title{The Hubble PanCET program: Long-term chromospheric evolution and flaring activity of the M dwarf host GJ 3470\thanks{Synthetic XUV spectra of GJ\,3470 associated with the quiescent phases and flaring episodes of the six epochs of observations (appendix Fig.~\ref{fig:EUV_spectra}) are available at the CDS via anonymous ftp to cdsarc.u-strasbg.fr (130.79.128.5) or via http://cdsarc.u-strasbg.fr/viz-bin/qcat?J/A+A/}}

\author[1]{V.~Bourrier}
\author[1]{L.~A.~dos Santos}
\author[2]{J.~Sanz-Forcada}
\author[7]{A.~Garc\'ia Mu{\~n}oz}
\author[6]{G.~W.~Henry}
\author[10]{P.~Lavvas}
\author[9]{A.~Lecavelier}
\author[11]{M.~L\'opez-Morales}
\author[8]{T.~Mikal-Evans}
\author[4,5]{D.~K.~Sing}
\author[3]{H.~R.~Wakeford}
\author[1]{D.~Ehrenreich}

\affil[1]{Observatoire Astronomique de l'Universit\'e de Gen\`eve, Chemin Pegasi 51b, CH-1290 Versoix, Switzerland}
\affil[2]{Centro de Astrobiologia (CSIC-INTA), ESAC Campus, P.O. Box 78, E-28691 Villanueva de la Canada, Madrid, Spain}
\affil[3]{School of Physics, University of Bristol, HH Wills Physics Laboratory, Tyndall Avenue, Bristol BS8 1TL, UK}
\affil[4]{Department of Earth \& Planetary Sciences, Johns Hopkins University, Baltimore, MD, USA}
\affil[5]{Department of Physics \& Astronomy, Johns Hopkins University, Baltimore, MD, USA}
\affil[6]{Center of Excellence in Information Systems, Tennessee State University, Nashville, TN 37209, USA}
\affil[7]{AIM, CEA, CNRS, Universit\'e Paris-Saclay, Universit\'e de Paris, F-91191 Gif-sur-Yvette, France} 
\affil[8]{Kavli Institute for Astrophysics and Space Research, Massachusetts Institute of Technology, Cambridge, MA, USA} 
\affil[9]{Sorbonne Universit\'e, CNRS, UMR 7095, Institut d’Astrophysique de Paris, 98bis bd Arago, 75014 Paris, France} 
\affil[10]{Groupe de Spectrom\'etrie Moleculaire et Atmosph\'erique, Universit\'e de Reims Champagne Ardenne, Reims, France}
\affil[11]{Center for Astrophysics, Harvard \& Smithsonian, 60 Garden Street, Cambridge, MA 01238, USA}

\authorrunning{V.~Bourrier et al.}
\titlerunning{Chromospheric behavior of the M dwarf GJ\,3470}

\offprints{V.B. (\email{vincent.bourrier@unige.ch})}

\institute{}

   \date{} %Received ...; accepted ...}
 
  \abstract
{

Neptune-size exoplanets seem particularly sensitive to atmospheric evaporation, making it essential to characterize the stellar high-energy radiation that drives this mechanism. This is particularly important with M dwarfs, which emit a large and variable fraction of their luminosity in the ultraviolet and can display strong flaring behavior. 

The warm Neptune GJ\,3470b, hosted by an M2 dwarf, was found to harbor a giant exosphere of neutral hydrogen thanks to three transits observed with the Hubble Space Telescope Imaging Spectrograph (HST/STIS). Here we report on three additional transit observations from the Panchromatic Comparative Exoplanet Treasury (PanCET) program, obtained with the HST Cosmic Origin Spectrograph (COS). These data confirm the absorption signature from GJ\,3470b's exosphere in the stellar Lyman-$\alpha$ line and demonstrate its stability over time. No planetary signatures are detected in other stellar lines, setting a 3$\sigma$ limit on GJ\,3470b's far-ultraviolet (FUV) radius at 1.3 times its Roche lobe radius. 

We detect three flares from GJ\,3470. They show different spectral energy distributions but peak consistently in the \ion{Si}{iii} line, which traces intermediate-temperature layers in the transition region. These layers appear to play a particular role in GJ\,3470's activity as emission lines that form at lower or higher temperatures than \ion{Si}{iii} evolved differently over the long term. Based on the measured emission lines, we derive synthetic X-ray and extreme-ultraviolet (X+EUV, or XUV) spectra for the six observed quiescent phases, covering one year, as well as for the three flaring episodes. Our results suggest that most of GJ\,3470's quiescent high-energy emission comes from the EUV domain, with flares amplifying the FUV emission more strongly. The neutral hydrogen photoionization lifetimes and mass loss derived for GJ\,3470b show little variation over the epochs, in agreement with the stability of the exosphere. 

Simulations informed by our XUV spectra are required to understand the atmospheric structure and evolution of GJ\,3470b and the role played by evaporation in the formation of the hot-Neptune desert.
}


   \maketitle

%%%%%%%%%%%%%%%%%%%%%%%%%%%%%%%%%%%%%%%%%%%%%%%%%%%%%%%%%%%%%%%%%%%%%%%%%%%%%%%%%

\section{Introduction}

High-energy stellar radiation plays an important role in the structure and chemistry of exoplanetary atmospheres and their evolution. X-ray and extreme ultraviolet (XUV) radiation was proposed as the source for the hydrodynamical expansion (e.g., \citealt{VM2003}; \citealt{Lammer2003}; \citealt{Lecav2004}; \citealt{GarciaMunoz2007}; \citealt{Johnstone2015}; \citealt{Guo2016}) that leads to the evaporation of close-in hot Jupiters (\citealt{VM2003,VM2004,VM2008}; \citealt{Ehrenreich2008}; \citealt{BJ_Hosseini2010}; \citealt{Lecav2010,Lecav2012}; \citealt{Bourrier2013,Bourrier2020_MOVESIII}). The structure of the close-in planet population (e.g., \citealt{Lecav2007}; \citealt{Davis2009}; \citealt{SanzForcada2010,SanzForcada2010corr}; \citealt{szabo_kiss2011}; \citealt{Mazeh2016}), direct observations (e.g., \citealt{Ehrenreich2015}; \citealt{Bourrier2018_GJ3470b}), and evolution simulations (e.g., \citealt{Owen2012}; \citealt{Lopez2013}; \citealt{Jin2014}; \citealt{Kurokawa2014}; \citealt{Owen2018}) all suggest that Neptune-size exoplanets are much more sensitive than hot Jupiters to atmospheric escape. Giant clouds of neutral hydrogen, in particular, have been observed around the warm Neptunes GJ\,436b (\citealt{Kulow2014}; \citealt{Ehrenreich2015}) and GJ\,3470b (\citealt{Bourrier2018_GJ3470b}). Yet it is not clear whether the evaporation of this class of planets stems from hydrodynamical escape or an intermediate regime with Jeans escape (\citealt{Bourrier2016}; \citealt{Salz2016a}; \citealt{Fossati2017}). Via hydrodynamical escape, heavy species are expected to be carried upward by collisions with the hydrogen outflow and to escape in substantial amounts (e.g., \citealt{VM2004}). Magnesium, silicon, and iron have been detected escaping from hot Jupiters (\citealt{VM2004}; \citealt{Linsky2010}; \citealt{Fossati2010, Fossati2013}; \citealt{Haswell2012}; \citealt{VM2013}, although see \citealt{Cubillos2020}; \citealt{Ballester2015}; \citealt{Sing2019}), supporting hydrodynamical escape as the source for their evaporation. For now though, no species heavier than hydrogen and helium have been observed in the upper atmospheres of warm Neptunes (\citealt{ParkeLoyd_2017}; \citealt{Lavie2017}; \citealt{DosSantos2019}; \citealt{Palle2020}; \citealt{Ninan2020}), leaving open the possibility that they are in an intermediate regime between hydrodynamical and Jeans escape, and that fractionation by mass occurs in the transition between the lower and upper atmosphere. Interestingly, both GJ\,436b and GJ\,3470b orbit M dwarf hosts, whereas the hot Jupiters around which atmospheric escape has been reported orbit earlier-type stars. The different irradiation associated with different spectral types might contribute to a variety of evaporation regimes. This highlights the importance of characterizing the irradiative environment of warm Neptunes around M dwarfs, whose ultraviolet activity remains poorly understood (e.g., \citealt{France2016}).\\

Spectral time series obtained with the Hubble Space Telescope (HST), in particular, provide an opportunity to observe and characterize far-ultraviolet (FUV) flares. Such events have previously been observed in G-, K-, and M-type stars  \citep{MitraKraev2005, Ayres2015_Acen, Loyd2018_MUSCLESV, Bourrier2020_MOVESIII}. Even quiet M dwarfs can display strong flaring behavior \citep[e.g.,][]{Paulson2006, France2012, Loyd2018_MUSCLESV, DosSantos2019}. \citet{Loyd2018_MUSCLESV} suggest that UV-only flares may be the result of reconnections in smaller magnetic structures that are only capable of heating the stellar atmosphere transition region and chromosphere, whereas larger structures may release enough energy to drive hotter X-ray emitting plasma up into the corona. Furthermore, \citet{Loyd2018_MUSCLESV} performed an extensive analysis of flares in active and inactive M dwarfs and found that, in general, their transition region flares occur at a rate $\sim$1000 times that of the Sun. According to \citet{France2012}, the quiet M4V dwarf GJ~876 can display FUV flares that are at least ten times stronger than the quiescent flux on timescales of hours to days. During flares, the transition region emission lines can exhibit a redshifted excess extending to 100 km~s$^{-1}$ in Doppler space, indicating a downflow of material toward the host star \citep{Loyd2018_MUSCLESV}. Such behavior is observed in flares from various types of stars, including the Sun, as the result of coronal and chromospheric condensation (e.g., \citealt{Fisher1989, Hawley2003, Bourrier2018_FUV}).\\

Close-in exoplanets can be susceptible to substantial changes in their upper atmosphere due to the high-energy activity of their host stars. \citet{Chadney2017} found that the neutral upper atmospheres of hot gas giants such as HD~209458~b and HD~189733~b are not significantly affected by flares from their G- and K-type hosts, which therefore do not significantly change the planetary atmospheric escape rates. On the other hand, \citet{Miguel2015} showed that the atmospheric photochemistry of Neptune-sized planets such as GJ~436b can change at less than the millibar level with variation in their Lyman-$\alpha$ irradiation. The largest changes concern the mixing ratios of H$_2$O and CO$_2$, which shield CH$_4$ molecules in the lower atmosphere from stellar irradiation. However, it is important to note that the Lyman-$\alpha$ emission line has a muted flare response when compared to the continuum and other FUV lines; in particular, \citet{Loyd2018_MUSCLESV} showed that the wings of the Lyman-$\alpha$ line, which is the dominant source of FUV flux in M dwarfs, exhibit an increase at least one order of magnitude lower than other FUV emission lines. In the case of Earth-sized rocky planets, \citet{France2020} has shown that the flaring activity (more specifically, the relative amount of time that the star remains in a flare state) is the controlling parameter for the stability of a secondary atmosphere after the first 5 Gyrs.	\\

In a previous study (\citealt{Bourrier2018_GJ3470b}) we reported on transits of GJ\,3470b observed with the HST in the FUV Lyman-$\alpha$ line. These observations, taken as part of the Panchromatic Comparative Exoplanet Treasury program (PanCET: GO 14767, P.I. Sing \& L\'opez-Morales), revealed the giant exosphere of neutral hydrogen surrounding the planet and allowed us to perform a first characterization of its high-energy environment. Here we complement these observations with PanCET transits of GJ\,3470b obtained with the HST in other FUV lines, allowing us to search for escaping species heavier than hydrogen and to refine our knowledge of GJ\,3470's chromospheric activity and XUV emission. The FUV transit observations and their reduction are presented in Sect.~\ref{sec:reduc}. In Sect.~\ref{sec:shortterm} we search for spectro-temporal variations in the spectral lines of GJ\,3470. The properties and evolution of the stellar quiescent emission are analyzed in Sect.~\ref{sec:quiet_em}, and the impact of GJ\,3470's variable high-energy spectrum on its planetary companion is discussed in Sect.~\ref{sec:impact}.

%%%%%%%%%%%%%%%%%%%%%%%%%%%%%%%%%%%%%%%%%%%%%%%%%%%%%%%%%%%%%%%%%%%%%%%%%%%%%%%%%
\section{Observations and data reduction}
\label{sec:reduc}

In \citet{Bourrier2018_GJ3470b} we performed a first study of GJ\,3470b and its host star in the ultraviolet, using the Space Telescope Imaging Spectrograph (STIS) instrument on board the HST. Three visits were scheduled around the planet transit on 28 November 2017 (Visit A), 4 December 2017 (Visit B), and 7 January 2018 (Visit C). Here we expand this analysis with later observations obtained with the Cosmic Origin Spectrograph (COS) instrument on board the HST. Three visits were again scheduled around the planet transit on 23 January 2018 (Visit D), 8 March 2018 (Visit E), and 23 December 2018 (Visit F). In each visit, two orbits were obtained before the transit, one during, and two after. Scientific exposures lasted 2703\,s in all HST orbits, except during the first orbit, which was shorter because of target acquisition (exposure time 1279\,s). Data obtained in time-tagged mode were divided into five sub-exposures for the first HST orbit of each visit (duration 255\,s) and into ten sub-exposures for all other orbits (duration 270\,s). 

For all visits, data were obtained in lifetime-position 4 on the COS detector, using the medium resolution G130M grating (spectral range 1125 to 1441\,\AA). The pixel size ranges from 2.1 to 2.7\,km\,s$^{-1}$, with a COS resolution element covering about seven pixels. Data reduction was performed using the CALCOS pipeline (version 3.3.5), which includes the flux and wavelength calibration. Uncertainties on the flux values are overestimated by the pipeline (see \citealt{Wilson2017,Bourrier2018_FUV}) and were defined using Eq. (1) in \citet{Wilson2017}. We further applied a reduction factor of 0.7 to the flux uncertainties, following an analysis of the temporal dispersion of the quiescent stellar emission (Sect.~\ref{sec:quiet_em}). For the sake of clarity, all spectra displayed in the paper are binned by three pixels, and their wavelengths tables are corrected for the radial velocity of GJ\,3470. As in previous studies using COS data (e.g., \citealt{Linsky2012,Bourrier2018_FUV}), we identified significant spectral shifts between the expected rest wavelength of the stellar lines and their actual position. Overall the shifts are consistent between visits but depend on the stellar line and its position on the detector, ranging from about -8\,km\,s$^{-1}$ for \ion{N}{v} to 4\,km\,s$^{-1}$ for \ion{O}{i}. Indeed, the shifts are a combination of biases in COS calibration (the most up-to-date wavelength solution provides an accuracy of $\sim$7.5\,km\,s$^{-1}$; \citealt{Plesha2018}) and the physical shifts of GJ\,3470 chromospheric emission lines, of which little is known for M dwarfs (\citealt{Linsky2012}). In an attempt to disentangle the two contributions, we compared the positions of GJ\,3470 chromospheric emission lines with those of the M dwarf GJ\,436 (COS/G130M data from \citealt{DosSantos2019}). We found a systematic offset of about 12\,km\,s$^{-1}$ between the spectra of the two stars and no clear similarities between their line shifts. Therefore, we did not attempt to correct the position of GJ\,3470 lines and, unless specified, all spectra hereafter are shown in the expected stellar rest frame.

We set the planetary system properties to the values used in Table 1 of  \citet{Bourrier2018_GJ3470b}. Rest wavelengths for the stellar lines were taken from the NIST Atomic Spectra Database (\citealt{Kramida2016}) and their formation temperatures from the Chianti v.7.0 database (\citealt{Dere1997,Landi2012}).

%%%%%%%%%%%%%%%%%%%%%%%%%%%%%%%%%%%%%%%%%%%%%%%%%%%%%%%%%%%%%%%%%%%%%%%%%%%%%%%%%
\section{Analysis of GJ\,3470 FUV lines in each epoch}
\label{sec:shortterm}

\begin{figure*}
\begin{minipage}[h!]{\textwidth}
\includegraphics[trim=0cm 0.5cm 0cm 0cm,clip=true,width=0.9\columnwidth]{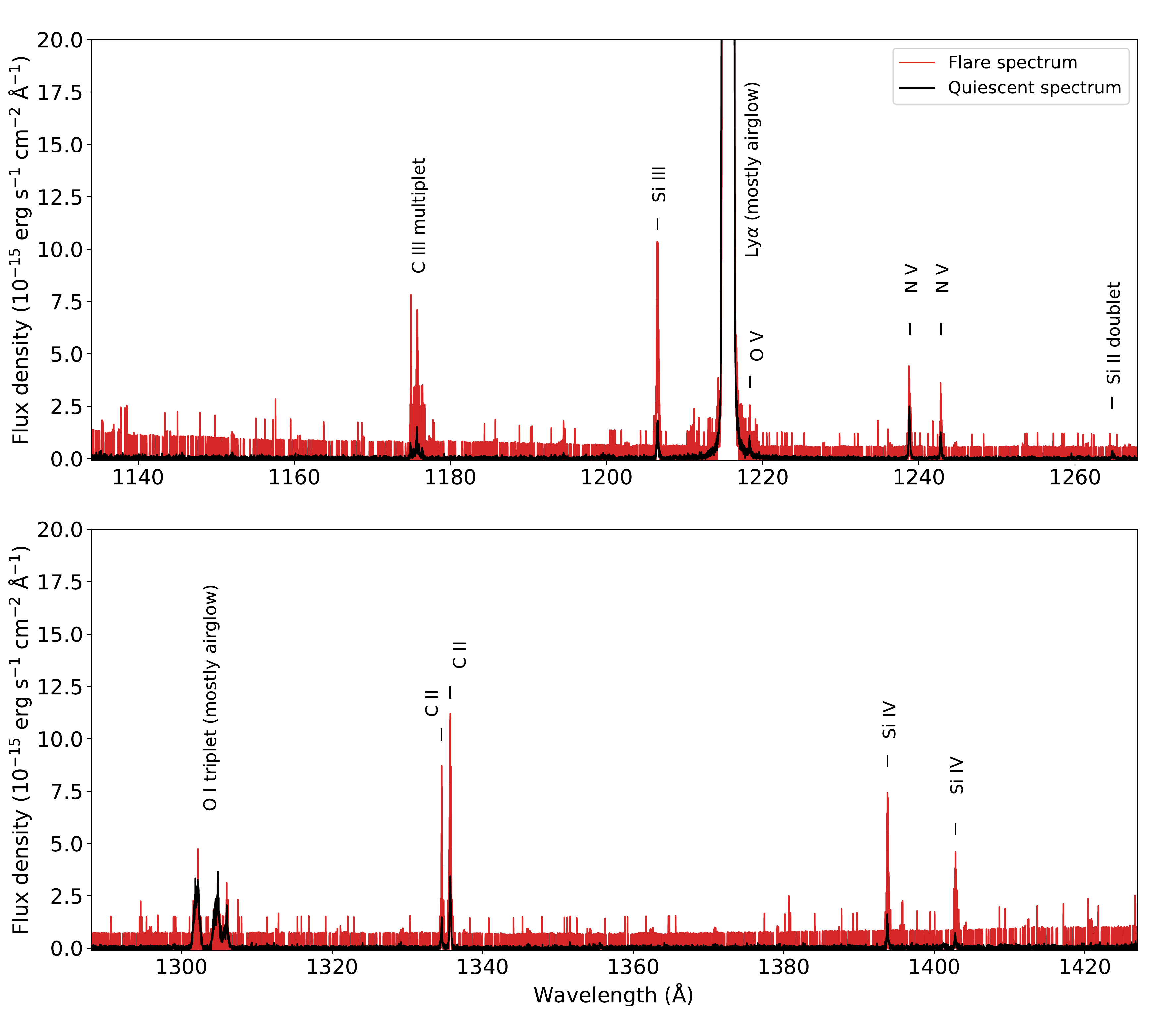}
\centering
\end{minipage}
\caption[]{COS spectrum of GJ\,3470, plotted as a function of wavelength in the star rest frame and averaged over Visits E and F. Flare brightenings are visible in most of the measured emission lines.}
\label{fig:global_spec.pdf}
\end{figure*}

The following emission lines from the chromosphere and transition region were found to be bright enough to be analyzed separatedly in each visit: the \ion{C}{iii} multiplet ($\lambda$1175); the \ion{Si}{iii} $\lambda$1206.5, Ly-$\alpha$ $\lambda$1215.7, \ion{O}{v} $\lambda$1218.3 lines; the \ion{N}{v} doublet ($\lambda$1238.8 and 1242.8); the ground-state \ion{O}{i} $\lambda$1302.2 and excited \ion{O}{i} $\lambda$1304.8 and \ion{O}{i} $\lambda$1306.0 lines; the ground-state \ion{C}{ii} $\lambda$1334.5 and excited \ion{C}{ii}$^{*}$ $\lambda$1335.7 line; and the \ion{Si}{iv} doublet ($\lambda$1393.8 and 1402.8). We followed the same approach as in \citet{Bourrier2018_FUV} to identify the quiescent stellar line profiles in each visit and the deviations from this quiescent emission that could be attributed to the planetary transit or short-term stellar activity (Fig.~\ref{fig:global_spec.pdf}). \\

One of the disadvantages of using the COS spectrograph instead of STIS for FUV observations is that it has a circular aperture instead of a slit, resulting in strong contamination of the Lyman-$\alpha$ (\ion{H}{I}) and \ion{O}{I} lines by airglow. However, the stellar emission lines that fall inside the contamination can still be recovered when the star is bright enough. Even for faint stars, such as the case of GJ~3470, this recovery is possible in exposures that are performed when the telescope is on the night side of the Earth. Following this approach we were able to recover the \ion{H}{I} and \ion{O}{I} lines of GJ~3470 in the COS spectra.

In the case of the Lyman-$\alpha$ line, the contamination is present even when the observations are performed on the night side of the Earth, but it is at least one order of magnitude weaker than on the day side. Thus, we needed to subtract the airglow emission using the same procedure detailed in \citet{Bourrier2018_FUV, DosSantos2019}. In summary, we divided each orbit into four subexposures and selected those with the lowest airglow contamination (see appendix Fig.~\ref{fig:Grid_Ly_airglow}). We then remove the contamination by fitting the amplitude and position of the observed airglow to a template\footnote{\footnotesize{Airglow templates for HST-COS are available at http://www.stsci.edu/hst/cos/calibration/airglow.html.}} and then subtracting the fitted feature. In practice, if the star is bright enough, the procedure can be performed for all (sub-)exposures without the need to select the cleanest ones. We note that the \ion{O}{v} line is superimposed on the red wing of the geocoronal Ly-$\alpha$ line, which was corrected for in each HST orbit using a second order polynomial to extract the stellar \ion{O}{v} line. 

The airglow contamination in the \ion{O}{I} lines is less dramatic than \ion{H}{I}; usually, only one quarter of the orbit is responsible for most of the contamination in the whole exposure. We were able to select the subexposures that displayed no contamination at all. We can discern the faint stellar emission from potentially low levels of airglow emission because the first is narrower than the latter.

%%%%%%%%%%%%%%%%%%%%%%%%%%%%%%%%%%%%%%%%%%%%%%%%%%%%%%%%%%%%%%%%%%%%%%%%%%%%%%%%%
\subsection{Flaring activity}

\begin{figure*}
\begin{minipage}[h!]{\textwidth}
\includegraphics[trim=0cm 0cm 0cm 0cm,clip=true,width=\columnwidth]{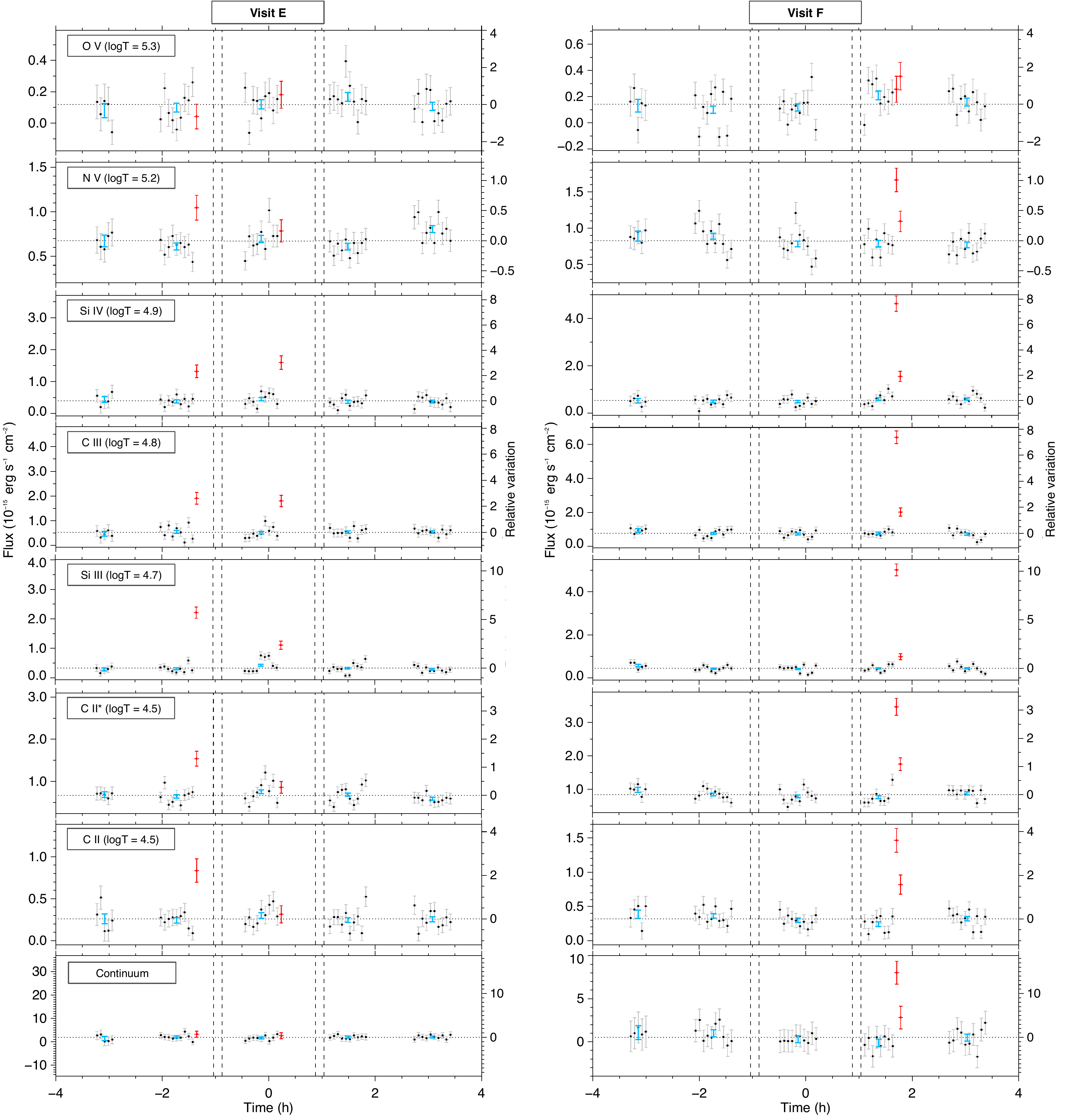}
\centering
\end{minipage}
\caption[]{Light curves of GJ\,3470 FUV lines in Visits E and F. Each row corresponds to a stellar line, ordered from top to bottom by decreasing formation temperature. The bottom row corresponds to the stellar continuum measured over the full ranges of the two detectors. Blue symbols correspond to the average flux over each HST orbit, flaring sub-exposures excluded (shown in red). For consistency, all fluxes are integrated over the full breadth of the lines, even though the flares may occur in specific spectral regions for some lines (see text). Vertical dashed lines are the transit contacts (there is no evidence for the planetary transit in any of the lines). Right axes indicate flux variations relative to the quiescent flux level (horizontal dotted lines) and are the same between visits for a given line. \ion{N}{v} and \ion{Si}{iv} fluxes are summed over the doublet lines.}
\label{fig:Flare_grid_LC}
\end{figure*}

Exploiting the temporal sampling of the data, we detected two independent flares in the last sub-exposures of the second and third orbits in Visit E and a single flare in the last two sub-exposures of the fourth orbit in Visit F (Fig.~\ref{fig:Flare_grid_LC}). In the latter case, we are able to distinguish between the peak of the flare and its decay phase. 

For all three flares the brightening is stronger in the \ion{Si}{iii} line (formation temperature log\,T = 4.7; up to 100\% in Visit F), followed by the \ion{C}{iii} and \ion{Si}{iv} lines (log\,T = 4.8--4.9; up to 80\% in Visit F) and then the \ion{C}{ii} (log\,T = 4.5; $\sim$0.5-4\%) and \ion{N}{v} (log\,T = 5.2, $\sim$0.5--1\%) lines. The flares are not detected in the \ion{O}{v} line (log\,T = 5.3), but the precision of the data would hide a variation at the level of \ion{N}{v} (Fig.~\ref{fig:Flare_grid_LC}). This suggests that most of the energy released by the observed flares comes from the intermediate-temperature regions of GJ\,3470 transition region where \ion{Si}{iii} is formed, with both low and high-energy tails as observed for other M dwarfs (e.g., GJ\,876 in \citealt{France2016}). At the observed FUV wavelengths, the continuum emission of an M dwarf such as GJ\,3470 is not detectable in individual pixels. Nonetheless we were able to detect the stellar continuum emission by summing the flux over the entire spectral ranges of the two COS/G130M detectors (from 1132.5 to 1270.0\,\AA\, and from 1288.0 to 1425.5\,\AA, excluding the stellar and airglow emission lines). While the flares are not detectable in Visit E continuum, we can clearly see the peak and decay phases of the flare in Visit F continuum (Fig.~\ref{fig:Flare_grid_LC}). Our flaring observations of GJ~3470 can be compared with those of the old (10 Gyr) M dwarf Barnard's Star, in which \citet{France2020} observed two FUV flares that have different spectral responses depending on where the lines are formed (chromosphere or transition region). For the chromospheric flare, \citet{France2020} found that the \ion{C}{II} flux increased by an average factor of 4.2, while the transition region flare increased the \ion{N}{V} flux by an average factor of 3.5, which is comparable to the flares we observed in GJ~3470.\\

\begin{center}
\begin{figure}
\includegraphics[trim=1.8cm 4.5cm 3.5cm 8cm,clip=true,width=\columnwidth]{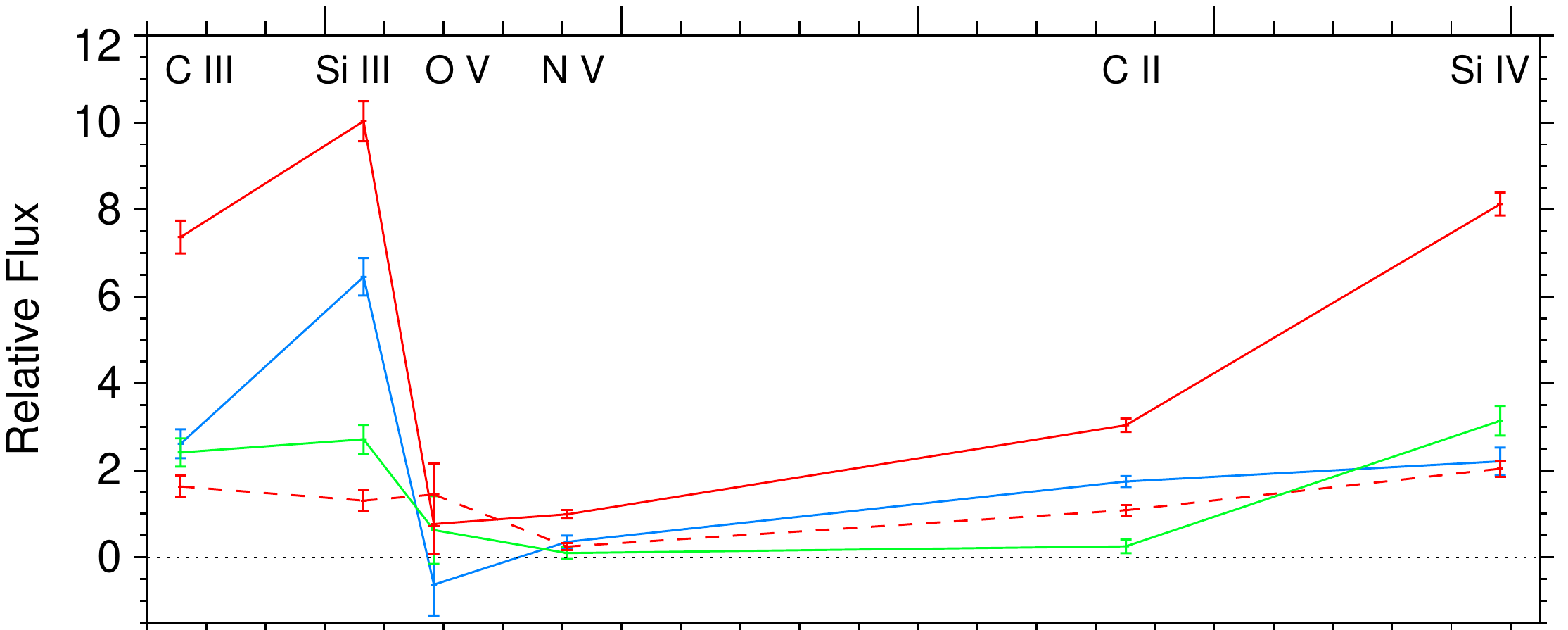}
\centering
\caption[]{Flaring spectra of GJ\,3470, relative to the quiescent stellar emission. Blue and green lines correspond to the two independent spectra measured in Visit E. Red corresponds to the flare measured in Visit F during the peak (solid line) and decay (dashed line) phases.}
\label{fig:LargeBand_Spec_vis}
\end{figure}
\end{center}

Fig.~\ref{fig:LargeBand_Spec_vis} highlights differences in the spectral energy distribution (SED) of the three flares. In Visit E the relative brightening of the \ion{Si}{iii} and \ion{C}{ii} lines is stronger in the first flare than in the second flare, while other lines show consistent variations. In Visit F the SED of the flare is also different from the two previous flares. Furthermore, we calculated the relative flux variations of the different lines between the peak and decay phases, and found that the \ion{Si}{iii} line decays more rapidly ($\sim$87\%) than lines formed at lower temperature (\ion{C}{ii}, $\sim$65\%) and higher temperatures (\ion{N}{v}, \ion{Si}{iv}, \ion{C}{iii}, $\sim$75\%). This traces the variable temporal response of GJ\,3470's chromosphere and transition region to its different flares and highlights the particular role played by the region where \ion{Si}{iii} is formed. Furthermore, one of the most defining characteristics of stellar flares is the strong increase in continuum fluxes at shorter wavelengths. This seems to be the case for Visit F, in which the continuum shows relative flux increases of 14.7$\pm$6.9 (peak) and 4.5$\pm$3.4 (decay) larger than for all measured flaring lines. A similar flare-to-quiescent continuum variation was observed in the flare star AD~Leo (\citealt{Hawley2003}), although the measured UV continuum wavelength ranges are different. We note that the two flares in Visit E again show a different behavior, with relative flux increases of 0.72$\pm$0.71 (first flare) and 0.39$\pm$0.70 (second flare) lower than for the strongest flaring lines. We checked that the differences between Visits E and F do not depend on the wavelength range over which the continuum is defined, that is, that there is no contribution from the flaring line wings in the measured continuum.

Appendix Figs.~\ref{fig:Flare_grid_spec_VE} and \ref{fig:Flare_grid_spec_VF} show the comparison between the spectra of the quiescent and flaring lines in Visits E and F, respectively. Overall the flares affect the full breadth of the stellar lines, but the red wing of some lines appear to brighten more intensely. This is especially visible for the \ion{C}{ii}, \ion{Si}{iii}, and \ion{Si}{iv} lines. Furthermore, the flux amplification seems lower in the core of these lines than in their wings. These spectral variations could trace the motion of the plasma along the stellar magnetic field lines during the flares.

\begin{center}
\begin{figure}
\includegraphics[trim=0cm 0cm 0cm 0cm,clip=true,width=\columnwidth]{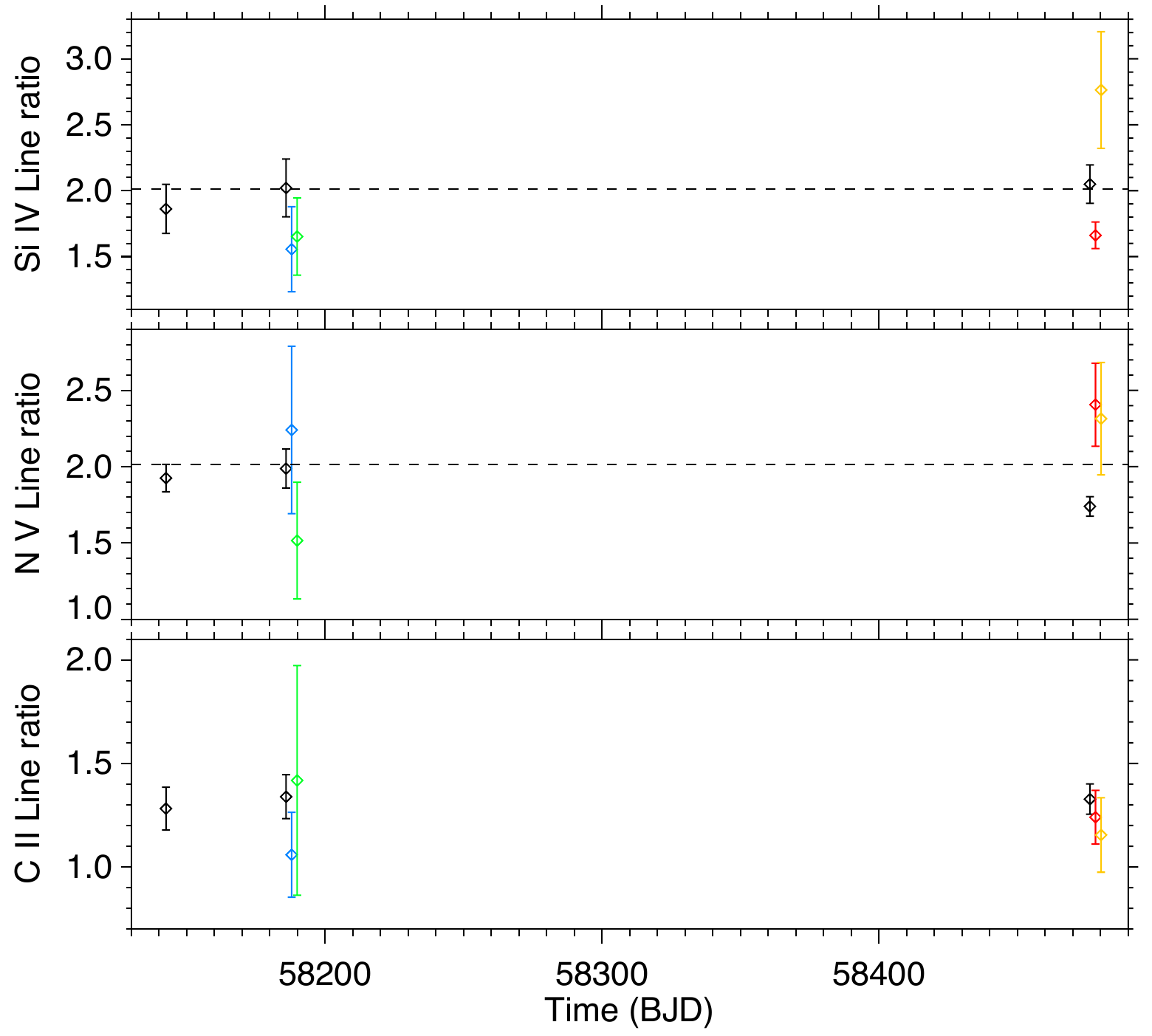}
\centering
\caption[]{Ratios of the \ion{Si}{iv} ($\lambda$1393.8 to 1402.8), \ion{N}{v} ($\lambda$1238.8 to 1242.8), and \ion{C}{ii} ($\lambda$1335.7 to $\lambda$1334.5) lines as a function of time in Visits D, E, and F. Black points correspond to the quiescent stellar emission. Blue and green points show the first and second flare in Visit E, while red and orange points show the peak and decay phase of the flare in Visit F, respectively. Horizontal dashed lines show the expected line ratios in an optically thin plasma for the \ion{Si}{iv} and \ion{N}{v} lines (the \ion{C}{ii} line ratio is at $\sim$2.5).}
\label{fig:LineRatios}
\end{figure}
\end{center}

We calculated the flux ratios of the \ion{Si}{iv}, \ion{N}{v}, and \ion{C}{ii} quiescent lines in Visit D, E, and F, and during the flaring sub-exposures (Fig.~\ref{fig:LineRatios}). We note that the \ion{C}{ii} lines were corrected for interstellar medium (ISM) absorption (Sect.~\ref{sec:ISM_GJ3470}). These flux ratios are sensitive to opacity effects in the stellar chromosphere and transition region and should be equal to about 2 for \ion{Si}{iv}, \ion{N}{v}, and 1.8 for \ion{C}{ii} in an optically thin plasma (\citealt{Pillitteri2015}). This is the case for the quiescent \ion{Si}{iv} and \ion{N}{v} emission, except for \ion{N}{v} in Visit F, which might be linked to the variability of the \ion{N}{v} $\lambda$1242.8 line in this visit (Sect.~\ref{sec:HE_lines}). The quiescent \ion{C}{ii} lines show a consistent ratio of $\sim$1.3, which indicates that they originate from an optically thick plasma (assuming that ISM absorption is not degenerate with the line amplitudes, \textit{i.e.}, that optically thick line profiles would remain well described by the Voigt profiles used to reconstruct the lines). We further observe changes in the plasma opacity during the flares. In Visit E both flares show \ion{Si}{iv} line ratios of about 1.6, with the other line measurements too imprecise to conclude to a departure from an optically thin plasma. In Visit F the \ion{Si}{iv} line ratio is again close to 1.6 during the peak phase but increases above 2.0 during the decay phase. In contrast, the \ion{N}{v} line ratios during peak and decay phases are larger than during quiescence, and the \ion{C}{ii} line ratios are marginally lower than during quiescence (Fig.~\ref{fig:LineRatios}).\\

%%%%%%%%%%%%%%%%%%%%%%%%%%%%%%%%%%%%%%%%%%%%%%%%%%%%%%%%%%%%%%%%%%%%%%%%%%%%%%%%%
\subsection{Flare metrics}

Rather unfortunately, the three observed flares occurred near the end of HST orbits. For the two flares in Visit E, we miss the decay phase that occured during Earth's occultations. Nevertheless, we observed part of the decay phase for the flare in Visit F, allowing us to fit a flare light curve model to the data and to estimate the flare duration, peak flux, absolute energy, and equivalent duration. We studied these metrics in the bandpass of each flaring line and in the broadband FUV$_{130}$ bandpass used by \citet{Loyd2018_MUSCLESV}. This encompasses most of the COS G130M range (minus geocoronal contamination and detector edges), and allows us to compare GJ3470 with other flaring stars.

\begin{center}
\begin{figure}
\includegraphics[trim=0cm 0cm 0cm 0cm,clip=true,width=\columnwidth]{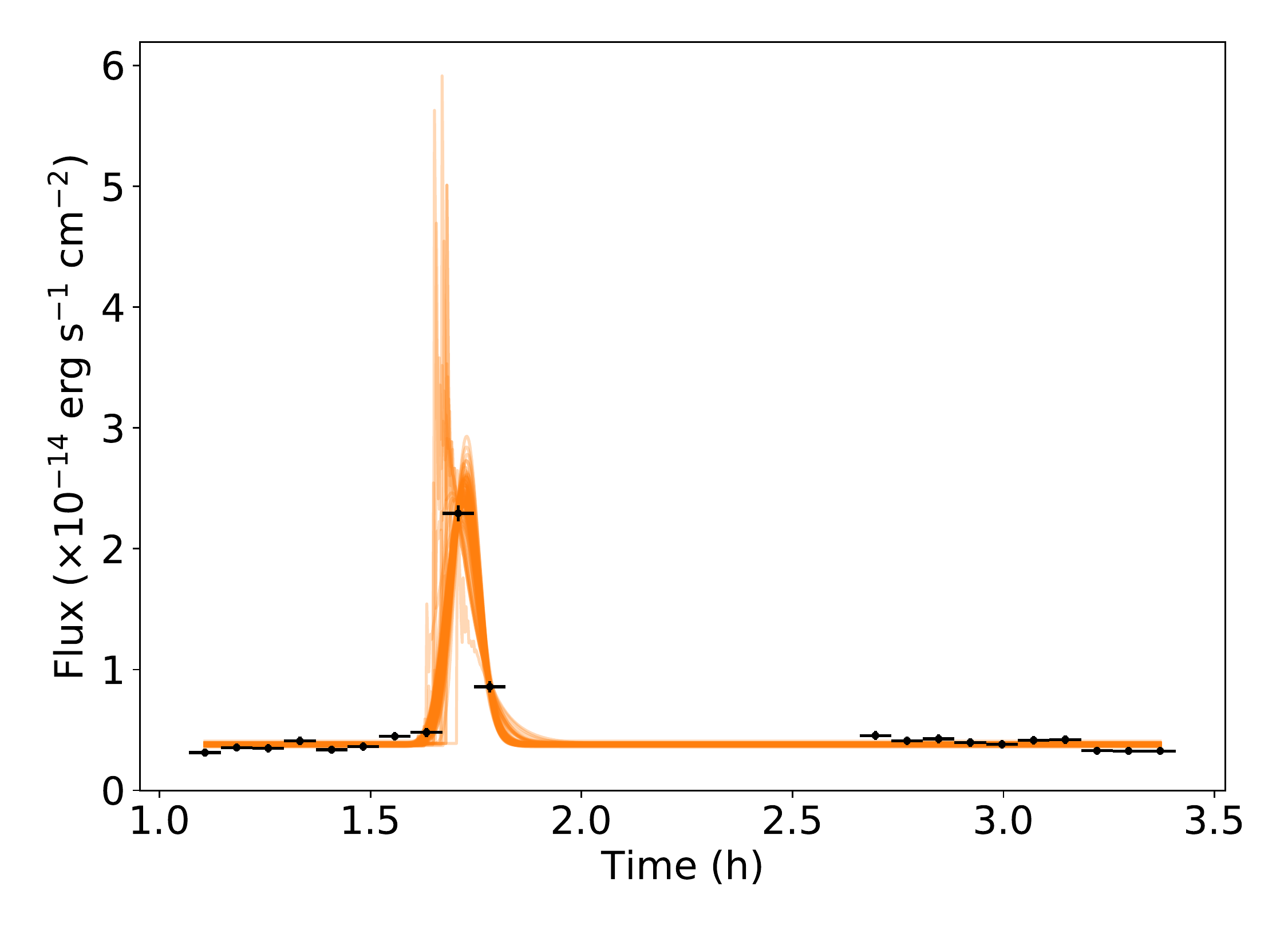}
\centering
\caption[]{Results of the light curve fit to the Visit F flare in the FUV$_{130}$ bandpass. The orange curves represent a random sample of 100 light curves drawn from the posterior distribution of the fit.}
\label{fig:model_LC}
\end{figure}
\end{center}

We modeled the flare light curve using the description of \citet{Gryciuk2016}, which was originally developed to model X-ray flare light curves of the Sun. This model assumes that a single-peaked flare can be described by two temporal profiles: one Gaussian energy deposition function that reflects the impulsive energy release and one exponential decay function representing the process of energy losses. The model has five free parameters: the amplitude of the flare, the time of peak flux, the rise time to reach peak flux, the decay time, and the baseline flux. We fit our observed light curves using \texttt{emcee} (\citealt{Foreman2013}) and then derive the flare metrics from the posterior samples (e.g., appendix Fig.~\ref{fig:corner_flare_F130}). Results are shown in Table \ref{table_flares}, with a sample of the best-fit models to the FUV$_{130}$ bandpass shown in Fig.~\ref{fig:model_LC}. 

\begin{table*}
\caption{Metrics for the flare observed in Visit F.}
\label{table_flares}
\centering
\begin{tabular}{l c c c c}
\hline
& Duration & Peak flux & Absolute energy & Equivalent duration \\
& (s) & ($\times 10^{-14}$ erg s$^{-1}$ cm$^{-2}$ \AA$^{-1}$) & ($\times 10^{27}$ erg) & (s) \\
\hline
FUV$_{130}$ passband & $832^{+90}_{-73}$ & $2.430^{+0.149}_{-0.102}$ & $689^{+32}_{-28}$ & $1760^{+100}_{-90}$ \\
\ion{Si}{III} & $604^{+90}_{-155}$ & $0.551^{+0.281}_{-0.052}$ & $144^{+32}_{-16}$ & $3440^{+790}_{-460}$ \\
\ion{Si}{IV} doublet & $547^{+212}_{-122}$ & $0.642^{+0.677}_{-0.147}$ & $154^{+77}_{-23}$ & $2700^{+1380}_{-500}$ \\
\ion{C}{III} multiplet & $547^{+196}_{-98}$ & $0.920^{+0.450}_{-0.233}$ & $209^{+64}_{-27}$ & $2830^{+860}_{-430}$ \\
\ion{C}{II} & $473^{+204}_{-106}$ & $0.214^{+0.352}_{-0.062}$ & $55^{+39}_{-11}$ & $2040^{+1410}_{-490}$ \\
\ion{C}{II}* & $832^{+131}_{-106}$ & $0.358^{+0.065}_{-0.030}$ & $115^{+16}_{-12}$ & $1410^{+240}_{-170}$ \\
\hline
\end{tabular}
\begin{tablenotes}[para,flushleft]
Notes: We were unable to fit the flare light curve model to the \ion{O}{v} and \ion{N}{v} lines because of their low brightening.
\end{tablenotes}
\end{table*}

The duration of the flare observed in Visit F suggests it is single-peaked and not complex, based on a comparison with the MUSCLES sample (\citealt{France2016,Loyd2018_MUSCLESV}). With an equivalent duration of $\sim$1760~s, and based on the power-law relationship in Section 3 of \citet{Loyd2018_MUSCLESV}, we infer that the cumulative rate of flares of the same absolute energy as or higher than that in Visit F is $1.73^{+1.50}_{-0.78}$ d$^{-1}$. We also used the \ion{Si}{iv} flux in the flare of Visit F to estimate the high-energy ($>$10\,MeV) proton fluence based on the scaling relations from \citet{Youngblood2017}. We estimate that this flare has a high-energy proton fluence of $54.8^{+110.6}_{-35.2}$ pfu\,s (1 pfu = 1 proton cm$^{-2}$ s$^{-1}$ sr$^{-1}$) at 1 au, which is consistent with an energetic (class-X) solar flare accompanied by a coronal mass ejection (\citealt{Cliver2012}). For comparison, flares of class X2-X3 occur approximately once a month in the Sun, and the old star GJ~699 (M3.5 type, 0.163 M$_\odot$) displays a rate of typical class-C to -X flares of approximately six per day (\citealt{France2020}). Furthermore, the equivalent durations of the flares of GJ~699 are approximately twice as long as those of GJ~3470 (\citealt{France2020}). These results suggest that even M dwarfs similarly classified as inactive can display significantly different high-energy activity, and that lower-mass M dwarfs exhibit relatively stronger flares even at old ages.

%%%%%%%%%%%%%%%%%%%%%%%%%%%%%%%%%%%%%%%%%%%%%%%%%%%%%%%%%%%%%%%%%%%%%%%%%%%%%%%%%
\subsection{Short-term variations}

\subsubsection{High-energy lines}
\label{sec:HE_lines}

Once flaring exposures were excluded, we compared in each visit the spectra averaged over each HST orbit to search for further short-term variations. Most lines are stable, but we describe hereafter several deviations from quiescent emission.

%\begin{itemize}
%\item 
In Visit E several sub-exposures in the second orbit show an increased flux in the \ion{Si}{iii} line. This occurs just before the first flare, but there is no evidence for a link (Fig.~\ref{fig:Flare_grid_LC}). 

\begin{center}
\begin{figure}
\includegraphics[trim=0cm 0cm 0cm 0cm,clip=true,width=\columnwidth]{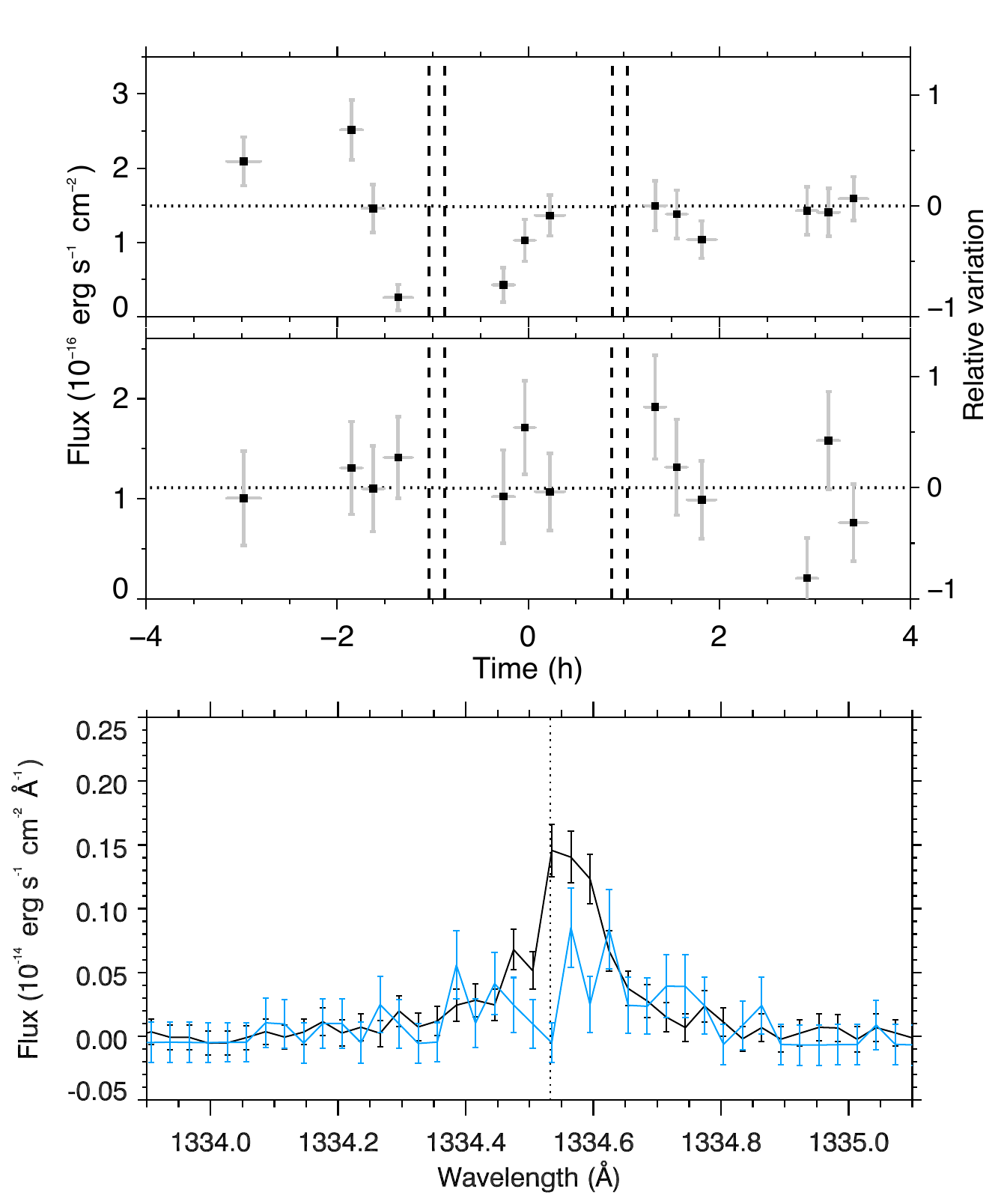}
\centering
\caption[]{\ion{C}{ii} $\lambda$1334.5 line in Visit D. \textit{Top panel:} Temporal evolution of the line flux averaged over the variable range -17 to 14\,km\,s$^{-1}$ (top) and over its stable complementary range (bottom). \textit{Bottom panel:} Comparison between the stellar line averaged over the first, last, and second orbit (black profile), and the line averaged over the two most absorbed sub-exposures visible in the top panel (blue profile). The wavelength scale is in the expected star rest frame and has not been corrected for the 3.6\,km\,s$^{-1}$ line shift (see text).}
\label{fig:Fig_CII_V1_paper}
\end{figure}
\end{center}

%\item
In Visit D the core of the \ion{C}{ii} $\lambda$1334.5 line dims dramatically from just before ingress to mid-transit. Taking the first and last two orbits as reference for the quiescent line, the flux decreases by up to 78$\pm$9\% over the velocity range -14 to 17\,km\,s$^{-1}$ (Fig.~\ref{fig:Fig_CII_V1_paper}). We note that this range is blueshifted by 3.6\,km\,s$^{-1}$ with respect to the line rest frame. Following this drop the flux actually remains lower than its pre-transit level, possibly suggesting that the planet is both preceded by a bow-shock and trailed by a tail of carbon atoms (see, e.g., \citealt{Bourrier2018_FUV}). We however lack data to attribute a planetary origin to these variations, which could arise from stellar variability as well. Indeed, the \ion{C}{ii}$^{*}$ $\lambda$1335.7 line displays no equivalent variation and it drops in flux in the last orbit, which favors a higher variability of the stellar \ion{C}{ii} lines in this visit.

%\item 
In Visit E the \ion{N}{v} $\lambda$1238.8  line shows significant variations in its red wing over the entire visit, while its blue wing remains stable. The \ion{N}{v} $\lambda$1242.8 line shows no equivalent variation, and its core brightens significantly in the last orbit of the visit. The absence of correlation between the two doublet lines or with the planet transit favors a stellar origin for these variations. 

%\item
In Visit F the \ion{N}{v} $\lambda$1238.8 line shows no significant variations, while the core of the \ion{N}{v} $\lambda$1242.8 line dims in the second orbit of the visit. Although this orbit is within the planet transit, the absence of correlation between the two doublet lines and their overall variability over the COS visits again favors a stellar origin.\\
%\end{itemize}

We then searched for the broadband FUV transit of GJ\,3470b by cumulating all measured lines in their quiescent states, excluding the \ion{H}{i} and \ion{O}{i} lines because of possible uncertainties in the airglow correction. Line fluxes in all sub-exposures were normalized by their out-transit values and averaged independently over the pre-transit, in-transit, and post-transit phases. Flaring and variable sub-exposures were excluded from this calculation. The average stellar flux was found to be stable outside of the transit (relative variation between the post and pre-transit phases of -2.9$\pm$4.1\%). We measure a relative flux variation of 3.3$\pm$3.8\% during the transit, which sets an upper limit on the FUV continuum radius of GJ\,3470b of 3.4 (1$\sigma$) and 4.7 (3$\sigma$) times its radius at 358\,nm (4.8$\pm$0.2$\,R_{\Earth}$, \citealt{Chen_2017}). For comparison, the Roche lobe of GJ\,3470b has  an equivalent volumetric radius of 3.6\,R$_\mathrm{p}$.

\subsubsection{Analysis of the Lyman-$\alpha$ line}

\begin{center}
\begin{figure}
\includegraphics[trim=0.cm 0.5cm 0cm 0.5cm,clip=true,width=\columnwidth]{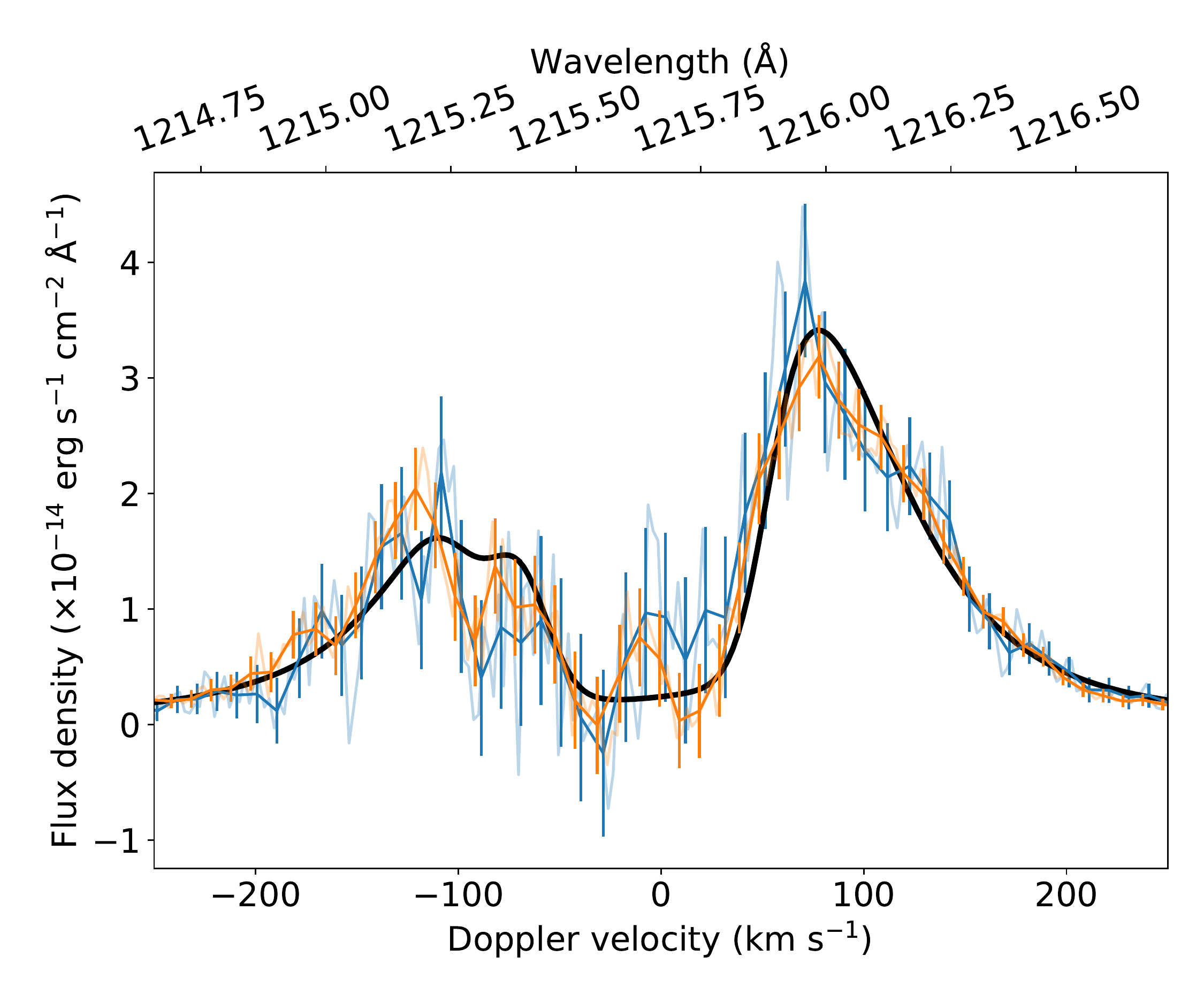}
\caption[]{Lyman-$\alpha$ line of GJ\,3470, plotted in the stellar rest frame. The blue and orange profiles show the COS spectra corrected for airglow contamination and averaged, respectively, over the out-of-transit and in-transit HST orbits in Visits D and E. Light profiles are the spectra at the original COS resolution, while strong profiles are binned with a 10\,km\,s$^{-1}$ resolution. The black profile is the best-fit model to the observed STIS Lyman-$\alpha$ line from \citealt{Bourrier2018_GJ3470b}.}
\centering
\label{fig:combined_lya_spectra}
\end{figure}
\end{center}

We successfully recovered the Lyman-$\alpha$ stellar emission in Visits D and E. Visit F yielded an anomalous decrease in the blue wing, possibly resulting from an airglow emission significantly different from the template in that region of the spectrum. It is unclear what could be the source of this mismatch between the observed airglow and the template. Thus we analyze only the Lyman-$\alpha$ time series in Visits D and E and discard Visit F. Fig.~\ref{fig:combined_lya_spectra} shows that there is an overall good agreement between the corrected COS Lyman-$\alpha$ spectra and the best-fit model derived by \citet{Bourrier2018_GJ3470b} from STIS spectra. This illustrates the stability of GJ\,3470 Lyman-$\alpha$ as well as the effectiveness of our airglow correction method.

\begin{center}
\begin{figure}
\includegraphics[trim=0.cm 0cm 0cm 0cm,clip=true,width=\columnwidth]{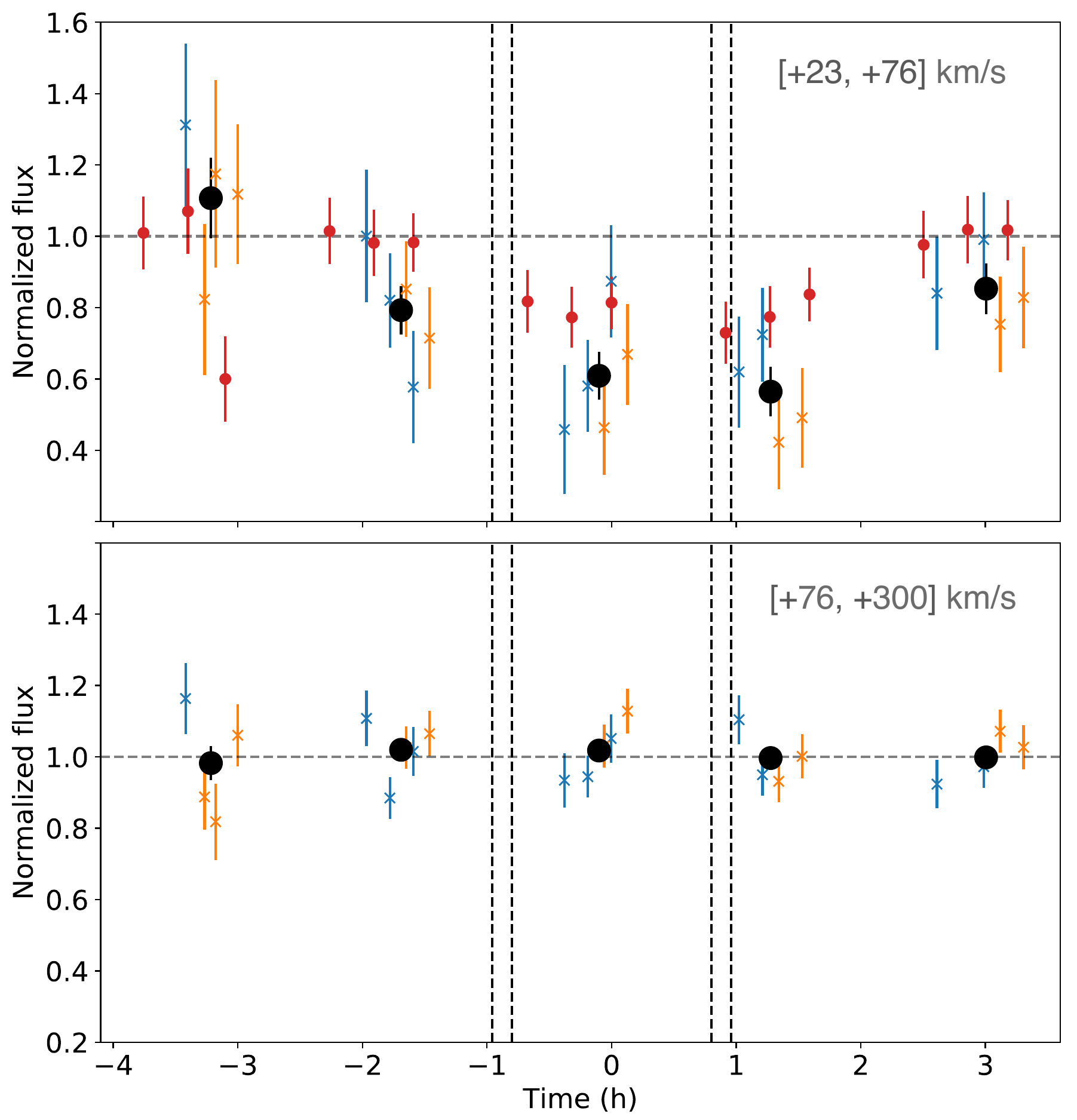}
\caption[]{Light curves of GJ\,3470 in the red wing of the Lyman-$\alpha$ line. The top and bottom panels show the flux integrated, respectively, over [+23, +76] and [+76, +300]\,km~s$^{-1}$ in the stellar rest frame. Blue and orange points correspond to COS exposures in Visits D and E, binned over the phase window of each HST orbit into the black points. Red points correspond to STIS exposures in Visits A, B, and C (\citealt{Bourrier2018_GJ3470b}). All data were normalized by the fluxes in the first and last HST orbits. Vertical dashed lines indicate the transit contacts.}
\centering
\label{fig:Lyalpha_LC}
\end{figure}
\end{center}

We computed the fluxes in the same ranges in which \citet{Bourrier2018_GJ3470b} found signals of planetary absorption (between Doppler velocities [-94, -41] and [+23, +76]\,km~s$^{-1}$ in the stellar rest frame), as well as in the stable ranges in the far wings of the stellar Lyman-$\alpha$ line (Fig.~\ref{fig:Lyalpha_LC}). Fluxes were normalized using the first and last orbits of each visit. Uncertainties are large in the blue wing as little stellar flux remains after geocoronal contamination correction. We measure absorptions of -6.8$\pm$16\% in Visit D and 62$\pm$55\% in Visit E, yielding an average value of 28$\pm$29\%. While consistent with the absorption measured in the STIS data set (35$\pm$7\%, \citealt{Bourrier2018_GJ3470b}), the large uncertainties prevent any meaningful comparison. On the other hand we were able to reproduce the red wing atmospheric signal in both Visits D and E (Fig.~\ref{fig:Lyalpha_LC}). The average decrease in the red wing flux that we observe in the COS data set is 39$\pm$7\%, which is marginally deeper than that observed by \citet[][23$\pm$5\% for the STIS data set]{Bourrier2018_GJ3470b}. The far red wing above Doppler velocity 76~km~s$^{-1}$ is stable up to 5\% (14\%) at 1$\sigma$ (3$\sigma$) confidence. The COS observations thus provide an independent confirmation of a neutral hydrogen cloud around GJ~3470~b and indicates that its redshifted absorption feature is stable across many planetary orbits.

%%%%%%%%%%%%%%%%%%%%%%%%%%%%%%%%%%%%%%%%%%%%%%%%%%%%%%%%%%%%%%%%%%%%%%%%%%%%%%%%%
\section{GJ3470 quiescent emission}
\label{sec:quiet_em}

To study the quiescent high-energy emission of GJ\,3470, we averaged the line fluxes over all orbits in a given visit, excluding those sub-exposures affected by flares and other short-term variations (Sect.~\ref{sec:shortterm}). The \ion{Si}{iii} and \ion{N}{v} line fluxes measured with STIS in Visits A, B, and C (\citealt{Bourrier2018_GJ3470b}) were included in this analysis. The integrated line fluxes measured with HST correspond to the intrinsic stellar flux, except for the \ion{C}{ii} doublet in which we identified the signature of the ISM.

\subsection{Interstellar medium toward GJ\,3470}
\label{sec:ISM_GJ3470}

\begin{center}
\begin{figure}
\includegraphics[trim=0.cm 0.5cm 1.cm 0.7cm,clip=true,width=\columnwidth]{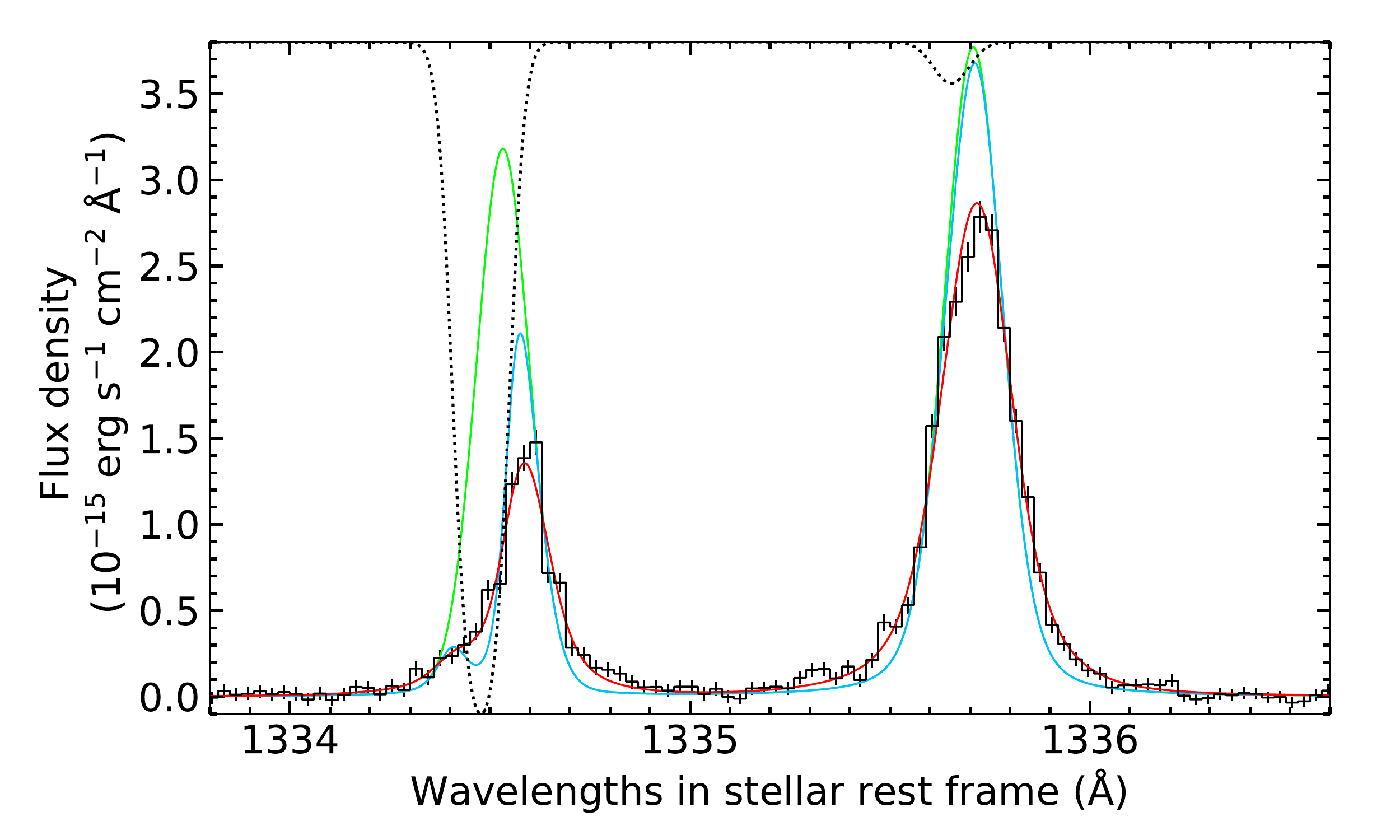}
\caption[]{Quiescent \ion{C}{ii} lines of GJ\,3470. Observed line profiles were corrected for their offsets with respect to their expected rest wavelength and averaged over Visits D, E, and F (black histogram). The green spectrum is the best fit for the intrinsic stellar lines at Earth's distance. It yields the blue profile after absorption by the ISM and the red profile after convolution with the COS LSF. The ISM absorption profile is plotted as a dotted black line (scaled to the panel vertical range).}
\centering
\label{fig:CII_doublet_fit}
\end{figure}
\end{center}

The ISM toward GJ\,3470 was studied by \citet{Bourrier2018_GJ3470b} using the STIS data, through its absorption of the stellar Lyman-$\alpha$ line in the \ion{H}{i} and \ion{D}{i} transitions. Due to the uncertainties introduced by the airglow correction in the COS data, we did not use the Lyman-$\alpha$ line to refine the ISM properties. Apart from this line, the signature of the ISM in the COS data is only detected in the stellar \ion{C}{ii} doublet, as expected from this lowly ionized, abundant ion. ISM absorption is clearly visible in the blue wing of the ground-state \ion{C}{ii}$\lambda$1334.5 line (Fig.~\ref{fig:CII_doublet_fit}), but marginally visible in the excited \ion{C}{ii}$^{*}$ $\lambda$1335.7 line. Following the same procedure as \citet{Bourrier2018_GJ3470b} with the Lyman-$\alpha$ line, we modeled both \ion{C}{ii} lines to refine the ISM properties toward GJ\,3470. We first measured the line shifts by fitting the wings of the \ion{C}{ii}$^{*}$ $\lambda$1335.7 line, unaffected by ISM absorption. Both lines were then aligned in the star rest frame and averaged over the three visits. 

The intrinsic stellar lines are better fitted with Voigt rather than Gaussian profiles. The LISM kinematic calculator\footnote{\mbox{\url{http://sredfield.web.wesleyan.edu/}}}, a dynamical model of the local ISM (\citealt{Redfield_Linsky2008}), predicts that the line of sight (LOS) toward GJ\,3470 crosses the LIC and Gem cloud. As a first approach we used a single-cloud model and fitted its temperature and independent column densities for the two \ion{C}{ii} transitions. Turbulence velocity has no impact on the fit and was set to 1.62\,km\,s$^{-1}$, which corresponds to either the LIC or the Gem cloud (\citealt{Redfield_Linsky2008}). The model spectrum was oversampled, convolved with the non-Gaussian and slightly off-center COS LSF\footnote{\mbox{\url{http://www.stsci.edu/hst/cos/performance/spectral_resolution/}}}, and resampled over the wavelength table of the spectra before comparison with the observations. The fit was carried out using the Markov chain Monte Carlo (MCMC) Python software package \textit{emcee} (\citealt{Foreman2013}).

The best-fit model is shown in Fig.~\ref{fig:CII_doublet_fit}. The asymetrical shape of the observed \ion{C}{ii}$\lambda$1334.5 line is well reproduced with a column density log$_{10}$\,$N_{\mathrm{ISM}}$(C\,{\sc ii})[cm$^{-2}$] = 14.12$\pm$0.05 and a radial velocity of -12.3$\pm$1.5\,km\,s$^{-1}$ for the ISM cloud relative to the star (we note this value is independent of the COS calibration bias). This value is in reasonably good agreement with the velocity predicted for the LIC (-8.2\,km\,s$^{-1}$) and with the value derived by \citet{Bourrier2018_GJ3470b} from the Lyman-$\alpha$ line (-7.7$\pm$1.5\,km\,s$^{-1}$). While the posterior probability distribution for log$_{10}$\,$N_{\mathrm{ISM}}$(C\,{\sc ii}$^{*}$)[cm$^{-2}$] shows a peak at about 12.7, it flattens below about 11.2 to a probability large enough that we cannot claim the detection of ISM absorption in the observed \ion{C}{ii}$^{*}$ $\lambda$1335.7 line (as hinted by its nearly symmetrical shape). Nonetheless, we can constrain log$_{10}$\,$N_{\mathrm{ISM}}$(C\,{\sc ii}$^{*}$)[cm$^{-2}$] $<$ 13.2 at 3$\sigma$, which sets an upper limit of about 0.12 on the density ratio between excited and ground-state carbon ions. The large temperature we derive from the line fits (1.4$\stackrel{+0.5}{_{-0.4}}\times$10$^{5}$\,K) supports the prediction that the LOS toward GJ\,3470 crosses two ISM clouds. However, there is no evidence for ISM absorption at the velocity predicted for the Gem cloud (6.9\,km\,s$^{-1}$), and a fit with a dual-cloud model did not allow us to disentangle their relative contributions.

\subsection{Long-term evolution}

\begin{center}
\begin{figure}
\includegraphics[trim=0cm 0cm 0cm 0cm,clip=true,width=\columnwidth]{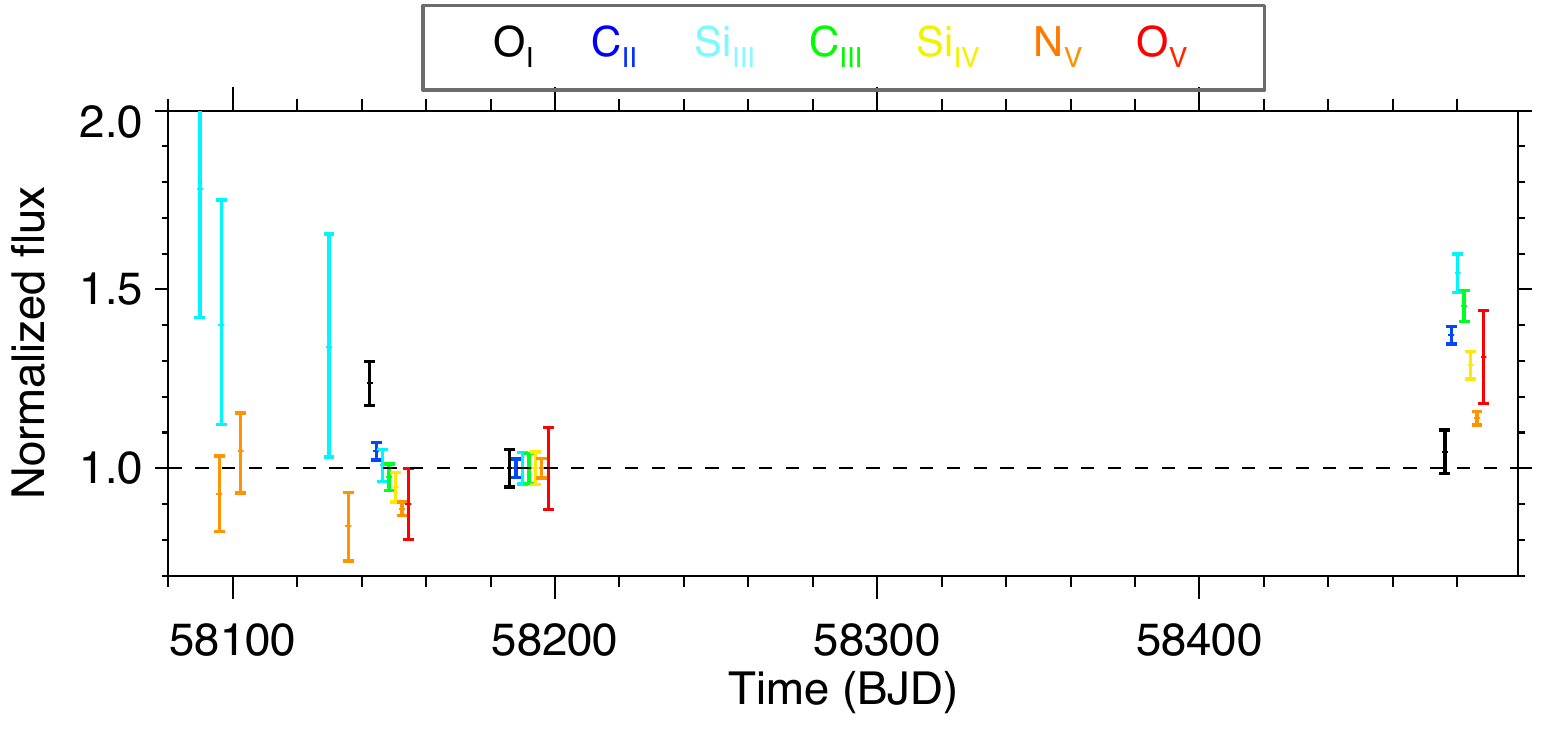}
\caption[]{Long-term evolution of GJ\,3470 quiescent FUV emission in the \ion{O}{i} triplet (black), \ion{C}{ii} doublet (blue), \ion{Si}{iii} line (cyan), \ion{C}{iii} multiplet (green), \ion{Si}{iv} doublet (yellow), \ion{N}{v} doublet (orange), and \ion{O}{v} line (red). Only the \ion{Si}{iii} and \ion{N}{v} lines could be measured with STIS in the first three epochs. In each visit, line values have been slightly offset in time by increasing formation temperature for the sake of clarity. }
\centering
\label{fig:TotF_vis}
\end{figure}
\end{center}

We show in Fig.~\ref{fig:TotF_vis} the evolution of GJ\,3470 FUV line fluxes from Visit A to F, relative to their flux in Visit D. The sensivitity of STIS does not allow measuring any significant variations of the \ion{Si}{iii} and \ion{N}{v} lines from Visits A to D, although the \ion{Si}{iii} line flux shows a tentative decrease in brightness over time. Most lines also show no significant variations over the $\sim$1.5 months separating Visits D and E, except for the lines formed at the lowest (\ion{O}{i}, log\,T = 3.9) and highest (\ion{N}{v}, log\,T = 5.2) temperatures. This is noteworthy because there is a hint that lines formed at low temperatures dimmed, while lines formed at high temperature brightened, their relative flux variation increasing as their formation temperature gets respectively lower or higher. The linchpin between dimming and brightening lines appear to be the \ion{Si}{iii} line (log\,T = 4.7). This behavior is clearer in the evolution from Visits E to F, separated by nearly nine months. All lines brightened between these two visits, but the relative flux amplification decreases as the line formation temperature gets away from that of \ion{Si}{iii}, which shows the strongest brightening. Interestingly, the \ion{O}{v} line seems to break away from this trend, with a larger flux variation than \ion{N}{i} despite a higher formation temperature (log\,T = 5.3). This suggests that the corona of GJ\,3470 evolves differently over time than its chromosphere.

This analysis shows a link between the long-term evolution of GJ\,3470 chromospheric structure and its short-term response to flares, with lines formed at lower or higher temperature than the \ion{Si}{iii} line displaying a different behavior.

\subsection{High-energy spectrum}
\label{sec:XUV_rec}

Among the measured lines, the \ion{N}{v} doublet presents a special interest as an activity tracer of the upper transition region. \citet{France2016} proposed that the \ion{N}{v} luminosity of K and M dwarfs may decline with rotation period. This trend was confirmed by \citet{France2018}, who showed that the UV activity of F-M stars decreases with rotation period as a power law, after an early saturated phase. The evolution of UV emission was investigated more specifically for early M dwarfs by \citet{Loyd2021}. Their empirical relation predicts an \ion{N}{v} surface flux of 6913$\stackrel{+1644}{_{-2528}}$erg\,cm$^{-2}$\,s$^{-1}$ for $P_{\mathrm{rot}}$ = 20.70$\pm$0.15\,d (\citealt{Biddle2014}), which is consistent with our measurement (4264\,erg\,cm$^{-2}$\,s$^{-1}$).  Furthermore, the relation from \citet{Loyd2021} yields an age of about 1.6\,Gyr for the measured \ion{N}{v} flux. Considering the uncertainties in this relation, this is in good agreement with the age of $\sim$2\,Gyr estimated by \citet{Bourrier2018_GJ3470b} from the Lyman-$\alpha$ flux and rotation period of GJ\,3470, further supporting the relatively young age of this star. \\

Most of the stellar EUV spectrum is absorbed by the ISM. We thus reconstructed synthetic XUV spectra of GJ\,3470 for the quiescent phases in each visit, as well as for the flaring episodes, using the emission measure distributions (EMD) from a coronal model calculated following \citet{Sanz-Forcada2011}. For the coronal region, the model is constrained by X-ray data of either GJ\,3470 alone or the combination of GJ\,3470 and AD Leo. For the transition region, the model is constrained by our measured FUV line fluxes. Compared to \citet{Bourrier2018_GJ3470b}, who used the \ion{Si}{iii} and \ion{N}{v} lines averaged over the STIS visits, we benefit here from the additional lines measured with high sensitivity in the COS data (\ion{C}{iii}, \ion{O}{v}, \ion{Si}{iv}, and \ion{C}{ii} corrected for the ISM in Sect.~\ref{sec:ISM_GJ3470}). Ten XUV spectra (six quiescent and four flaring) corresponding to the combined GJ\,3470 and AD Leo case, and generated in the region 1-1600\,\AA\,, are available in electronic form at the CDS (the EUV region is shown in Fig.~\ref{fig:EUV_spectra}, and the time evolution of EUV and FUV fluxes is shown in Fig.~\ref{fig:fluxevolution}). 

\subsubsection{EMD based on GJ\,3470 alone}
\label{sec:EMD_GJ3470}

\begin{center}
\begin{figure}
\includegraphics[trim=2.cm 1cm 3.5cm 3cm,clip=true,width=\columnwidth]{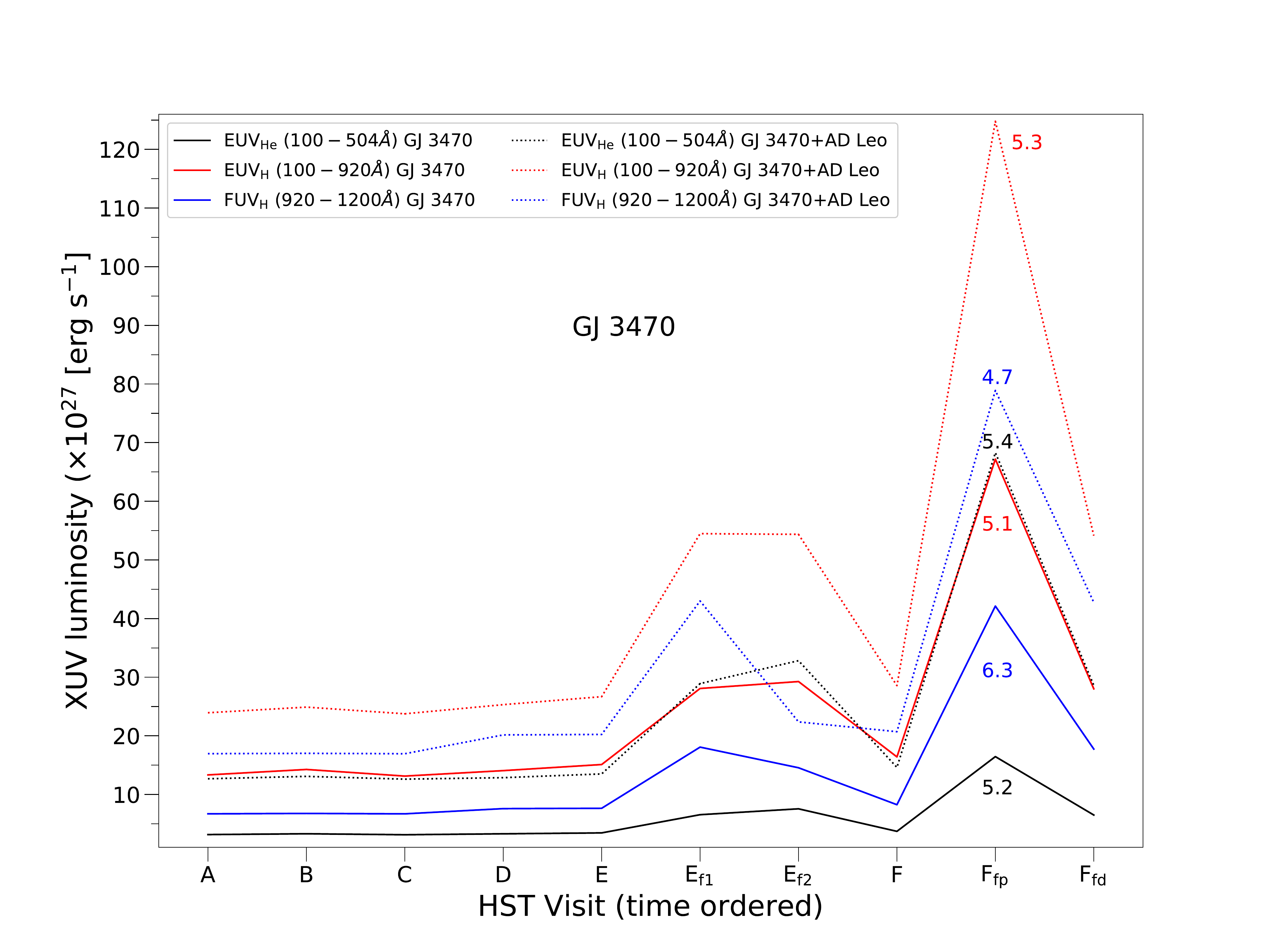}
\caption[]{Synthetic fluxes at 1\,au from GJ\,3470 in the FUV and two EUV bands as a function of time. We distinguish the fluxes calculated with the EMD model of GJ\,3470 alone (solid lines) and the EMD model of the combined GJ\,3470 and AD~Leo (dotted lines). Numbers over the flare peaks indicate the maximum-to-minimum ratio of the curves in each band. We note that variations in X-rays (5--100~\AA, not shown here) are limited to two main values in the case of the combined EMD ($5.7\times10^{27}$\,erg\,s$^{-1}$, quiescent; $1.8\times10^{28}$\,erg\,s$^{-1}$, flares), while it is almost constant for all intervals in the EMD of GJ 3470 alone ($2.3\times10^{27}$).} 
\centering
\label{fig:fluxevolution}
\end{figure}
\end{center}

We first constrained the coronal model with X-ray spectra of GJ\,3470 acquired on 2015/04/15 and obtained from the XMM-Newton archive (obs. ID 763460201, P.I. Salz). The model-derived quiescent fluxes display a slight increase over time in the EUV (100-920\,\AA) and FUV (920-1200\,\AA) bands, but remain stable overall (Fig.~\ref{fig:fluxevolution}). The XUV emission of GJ\,3470 is dominated by the EUV contribution, with an average value of 4.90\,erg\,cm$^{-2}$\,s$^{-1}$ over all epochs. This is on the same order as the Lyman-$\alpha$ flux (3.64\,erg\,cm$^{-2}$\,s$^{-1}$, \citealt{Bourrier2018_GJ3470b}), highlighting the importance of this line when considering the UV energy budget of a planet orbiting an M dwarf (e.g., \citealt{Youngblood2017}). 

The model-derived flare spectra show larger flux increases relative to the quiescent emission (Fig.~\ref{fig:fluxevolution}) in the FUV domain ($\sim$90-140\%, up to 410\% in Visit F) than in the EUV domain ($\sim$70-95\%, up to 310\% in Visit F). We used the same X-ray spectra to characterize quiescent and flaring stages (appendix Fig.~\ref{fig:emdmixcomp}). Thus we did not introduce any changes in the model at coronal temperatures, characterized by the X-ray spectral information. As a consequence, the modeled X-ray emission show little variation during the flares ($<$1\%), and the EUV emission is likely suffering smaller changes that we would have noticed if we had observed the coronal counterpart of the flares.\\

We found the EMD of GJ 3470 to be more sensitive to the flaring activity at a temperature of $\log T$(K)$\sim4.9$. To illustrate this we show the average of the $\log EM$(cm$^{-3}$) values in the temperature range $\log T$(K)$=4.7-5.1$ in Fig.~\ref{fig:mloss_phlife}. The EUV flux capable of ionizing H ($\lambda<920 \AA$) and He ($\lambda<504 \AA$) atoms is increased during flares (Fig.~\ref{fig:fluxevolution}), following the same trend observed in the Emission Measure, thus yielding larger mass-loss rates during flares (Fig.~\ref{fig:mloss_phlife}). The FUV flux differs from the EUV only in the second flare of Visit E, apparently less affected at these wavelengths.

\subsubsection{EMD based on GJ\,3470 and AD Leo combined}

As explained in Sect.~\ref{sec:EMD_GJ3470}, the available X-ray data for GJ\,3470 do not allow a separate coronal model for the quiescent and flaring stages. Although the modeled SED reflects the differences due to the UV emission during flares, we also need to understand how the flares are affecting the SED at temperatures above 1 MK, information given by X-ray (or EUV if available) spectra. To get that information, we used the dM3 star AD~Leo as a proxy, using the coronal models of quiescent and flaring stages calculated in \citet{SanzForcada2002}. The emission measure values of AD~Leo need to be normalized to the level of GJ~3470. We analyzed the XMM-Newton observation available in the archive for AD~Leo (observed on 14 May 2001, obs. ID 0111440101, PI Brinkman) and compared the results of the spectral fit of EPIC with those of GJ~3470. The GJ~3470 EPIC spectrum is fit with a 2-T model: $\log
T$(K)$=6.60^{+0.07}_{-0.14}$, $7.09^{+0.81}_{-0.25}$, and $\log EM$(cm$^{-3}$)$=49.92^{+0.07}_{-0.32}$, $49.29^{+0.28}_{-0.58}$, respectively, for an ISM absorption of $N_{\rm H}$(cm$^{-3}$)$=3\times10^{18}$ and [Fe/H]$=-0.30$. The AD~Leo EPIC spectrum is fit with a 3-T model: $\log T$(K)$=6.47\pm0.01$, $6.85\pm0.01$, $7.46\pm0.03$ and $\log EM$(cm$^{-3}$)$=50.84^{+0.04}_{-0.02}$, $51.07\pm0.02$, $50.69^{+0.02}_{-0.03}$, respectively, for an ISM absorption of $N_{\rm H}$(cm$^{-3}$)$=3\times10^{18}$ and [Fe/H]$=-0.27$. The XMM-Newton light curve of AD~Leo shows only small flares (flux increased by less than a factor of $\sim2$), similar to the GJ~3470 XMM-Newton light curve. The full procedure is described next:
  (i) Compare the total EM of both stars ($\log EM$(cm$^{-3}$)=51.37 for AD~Leo, 50.01 for GJ~3470. Considering also a difference of 0.03 dex in [Fe/H], we need to apply a shift of -1.4~dex to AD~Leo's EMD; (ii) Apply the shift to either the flaring and the quiescent EMD of AD~Leo, limited to the range $\log T$(K)$=6.0-7.4$, extending the EM values of the GJ~3470 EMD in the different intervals up to $\log T$(K)=6.0 to make them consistent with the AD~Leo shifted EMD at $\log T$(K)$\geqslant 6.0$ (Fig.~\ref{fig:emdmixcomp}). The flaring AD~Leo model is applied to the flaring intervals of the HST observations of GJ~3470, while the quiescent AD Leo~model is applied to the rest of intervals; (iii) Produce new SED for each interval using the combined coronal models (Fig.~\ref{fig:EUV_spectra}).

The fluxes calculated in this way show slightly higher values in the quiescent stages than the model based on just the GJ~3470 quiescent data. This is likely due to a better temperature coverage above $\log T$(K)$=5.7$, so it gives a better idea of the likely flaring behavior of GJ~3470 at high temperatures. The main effects of this procedure are noticed at the shortest wavelengths (X-rays), as can be seen in Fig.~\ref{fig:EUV_spectra}. It is also evident in the substantial increase in XUV flux during flares (Fig.~\ref{fig:fluxevolution}), as in the case of the EMD built using the GJ\,3470 data alone.

%%%%%%%%%%%%%%%%%%%%%%%%%%%%%%%%%%%%%%%%%%%%%%%%%%%%%%%%%%%%%%%%%%%%%%%%%%%

\section{Impact on the planetary atmosphere}
\label{sec:impact}

\begin{center}
\begin{figure}
\includegraphics[trim=0cm 0cm 0cm 0cm,clip=true,width=\columnwidth]{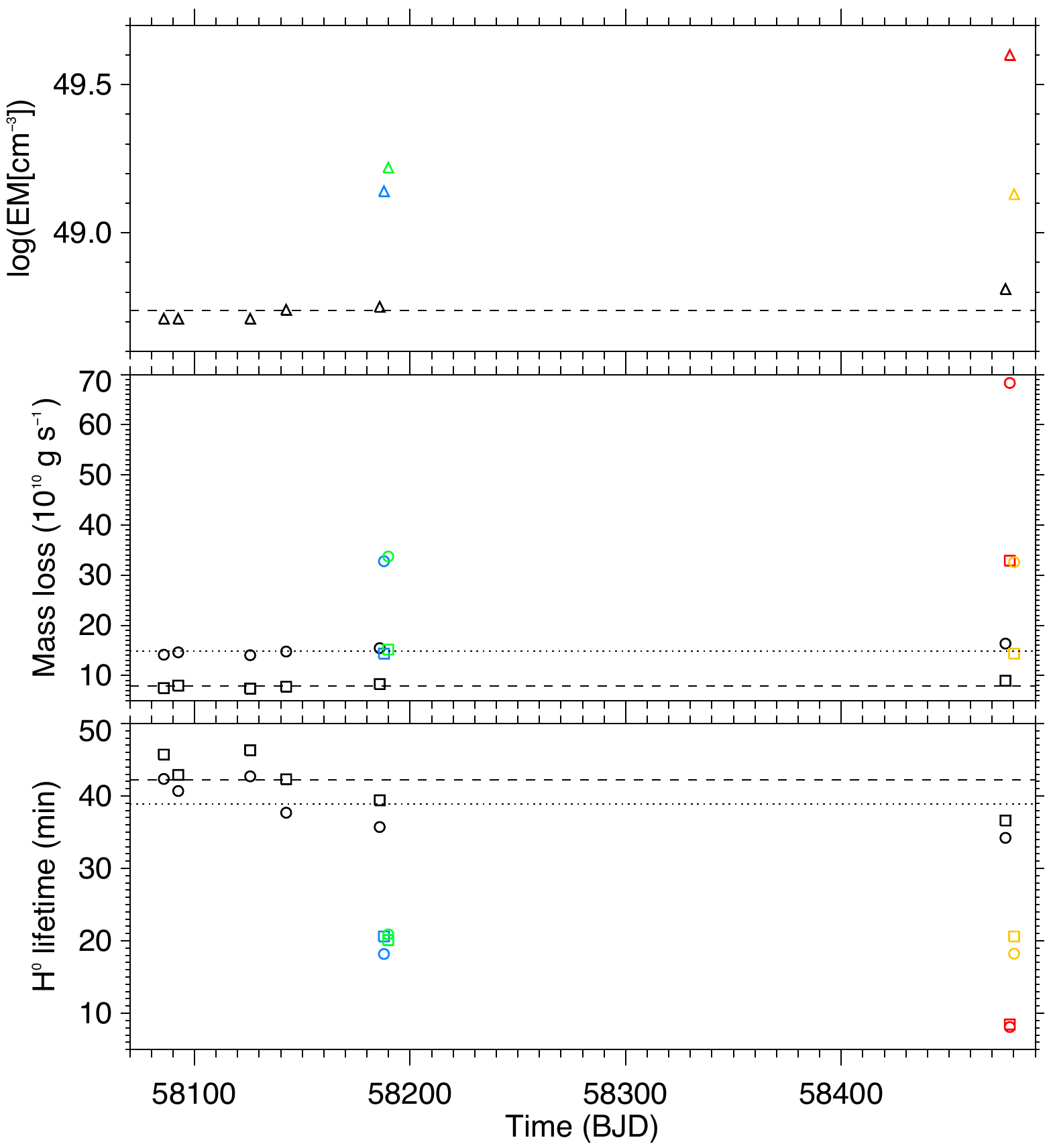}
\caption[]{Average emission measures ($\log EM$(cm$^{-3}$)) in the range $\log T$(K)$=4.7-5.1$ (upper panel), maximum energy-limited mass loss (middle panel), and photo-ionization lifetime of neutral hydrogen (lower panel) as a function of time. Black points correspond to the quiescent stellar emission. Blue and green points show the first and second flare in Visit E, while red and orange points show the peak and decay phase of the flare in Visit F, respectively. Squares and circles correspond, respectively, to the EMD based on GJ\,3470 alone and the combined GJ\,3470 and AD~Leo (emission measures in the plotted temperature range are the same in both cases). Horizontal and dotted dashed lines show the respective averages over quiescent values.}
\centering
\label{fig:mloss_phlife}
\end{figure}
\end{center}

The derived XUV spectra were used to estimate the conditions in GJ\,3470b's upper atmosphere during quiescent and flaring epochs (Fig.~\ref{fig:mloss_phlife}). We calculated the maximum total atmospheric mass-loss rate in the energy-limited regime (e.g., \citealt{Watson1981,Lammer2003,Erkaev2007,Lecav2007}), using the flux between 5 and 911.8\,\AA\, and assuming that all the energy input is used for atmospheric heating. Maximum mass loss remains stable over quiescent epochs, with an average value of about 8$\times$10$^{10}$\,g\,s$^{-1}$ (for the EMD based on GJ\,3470 alone) or 15$\times$10$^{10}$\,g\,s$^{-1}$ (for the EMD based on GJ\,3470 and AD~Leo). The larger mass loss in the combined GJ\,3470 + AD~Leo case is due to the larger stellar emission at short wavelengths. The energy released by the flares would then have increased maximum mass loss by up to $\sim$120\% in Visit E and by up to 320\% at the peak in Visit F. We note that numerical simulations suggest atmospheric mass loss may not be sensitive to typical flares as the enhanced energy input does not increase the outflow fluxes in proportion (\citealt{Chadney2017,Odert2020}), and the response time of the upper atmosphere may be longer than the flare duration (\citealt{Bisikalo2018}). Such conclusions, however, were obtained for hot Jupiters around G- or K-type stars and may not be applicable to Neptune-size planets around M dwarfs (\citealt{Atri2020}), as suggested by their different evaporation regime (e.g., \citealt{Owen2012,Bourrier2016,Bourrier2018_GJ3470b}).

We also calculated photo-ionization lifetimes for neutral hydrogen atoms at the semi-major axis of GJ\,3470b, following \citet{Bourrier2017_HD976}. Lifetimes overall decreased over the year covering our observations, in line with the increase in stellar XUV emission (Sect.~\ref{sec:XUV_rec}), but remained close to their average ($\sim$42\,min in the GJ\,3470 case, 39\,min in the combined case). The small difference between the two cases is due to the AD~Leo model not affecting the cool region of the EMD, which is responsible for the EUV flux photo-ionizing hydrogen atoms. Flaring episodes see photo-ionization lifetimes reduced by half, down to only $\sim$10\,min during the flaring peak in Visit F. Even if the atmospheric outflow was not substantially affected by the flares, such bursts of photons would have changed the ionization structure of the hydrogen exosphere and muted its Lyman-$\alpha$ transit signature.

Overall, the long-term stability of the mass loss and of the hydrogen photo-ionization lifetimes estimated above is consistent with the stability of GJ\,3470b's neutral hydrogen exosphere demonstrated by the repeatability of its Lyman-$\alpha$ transit signature (\citealt{Bourrier2018_GJ3470b}).\\

\section{Conclusion}
\label{sec:conclu}

GJ\,3470b is the second warm Neptune found to be evaporating, after GJ\,436b. Both planets orbit early M dwarfs and are located at the edge of the hot Neptune desert but display radically different exospheric structures. Characterizing the details of the radiative environment of such planets is thus essential to understanding how it shapes their atmospheric outflow. \\

STIS observations of GJ\,3470 in the FUV allowed probing the \ion{Si}{iii}, Ly-$\alpha$, and \ion{N}{v} stellar emission lines. The low sensitivity of the STIS/G140M grating, however, limited the information that could be derived on the XUV spectrum of GJ\,3470 and its temporal variability. Three additional transit data sets were thus obtained at later epochs with the COS spectrograph, providing access with high sensitivity to the \ion{C}{iii}, \ion{Si}{iii}, Ly-$\alpha$, \ion{O}{v}, \ion{N}{v}, \ion{O}{i}, \ion{C}{ii}, and \ion{Si}{iv} lines.

Analysis of the COS data at short cadence ($\sim$4.5\,min) revealed two flares of equivalent energy in Visit E and a stronger flare in Visit F, for which we could distinguish between the peak and decay phases. The measured emission lines form at different temperatures and thus probe different regions of the stellar atmosphere. We find that most of the energy released by the three flares comes from the intermediate-temperature regions ($\log$ T$\sim$4.7) of GJ\,3470 atmosphere where the \ion{Si}{iii} line is formed, with the other lines tracing low- and high-energy tails. The chromosphere of GJ\,3470 nonetheless displays a variable response to each flare, with different line brightenings and changes in plasma density relative to the quiescent state. Fitting the flare light curve in Visit F allowed us to study several metrics, which place GJ\,3470 in the inactive M dwarf branch (flare frequency of $\sim$1.7 per day) yet capable of emitting flares as energetic as more active stars. This suggests that even M dwarfs classified as inactive can display significant high-energy activity, showing that caution should be exercised when investigating the possible planetary origin of radio signals (e.g., \citealt{Mahadevan2021}). Besides the flares, we identify significant variations in the \ion{C}{ii} $\lambda$1334.5 (Visit D) and \ion{N}{v} $\lambda$1238.8 lines (Visits E and F), which most likely arise from stellar activity since no equivalent variations are measured in the second lines of these doublets.\\

The UV emission of GJ\,3470 evolved over the year covered by the STIS and COS visits, overall increasing over the last visits. As with the flares, chromospheric and transition region lines formed at lower and higher temperatures than the \ion{Si}{iii} line evolved differently, highlighting the pivotal role played by its formation region. Most of the stellar EUV spectrum, as well as specific FUV stellar lines, are absorbed by the ISM. Its properties toward GJ\,3470 were constrained by \citet{Bourrier2018_GJ3470b} from the reconstruction of the Lyman-$\alpha$ line. Here we detect its signature in the stellar \ion{C}{ii} doublet, whose reconstruction is consistent with the expected presence of at least two ISM clouds along the LOS. The reconstructed and measured stellar intrinsic fluxes are used to constrain a coronal model and derive the full XUV spectrum of GJ\,3470 in ten relevant epochs (six for the quiescent phases in each visit, two for Visit E flares, and two for the peak and decay phases of Visit F flare). Based on these model spectra, available at the CDS, we find that most of GJ\,3470 quiescent UV luminosity comes from the EUV domain while it is the FUV domain that displays the strongest relative brightening during flares. The X-ray domain contributes little to the overall stellar XUV emission and shows negligible variations from the flares. \\

We were able to correct the stellar Lyman-$\alpha$ and \ion{O}{i} lines from COS strong geocoronal contamination and to retrieve in Visits D and E the redshifted absorption signal of GJ\,3470b neutral hydrogen exosphere detected in the STIS data. While marginally deeper in the COS data (39$\pm$7\% vs 23$\pm$5\%), which could be due to uncertainties in the airglow correction, the exospheric transit light curve has the same shape in the STIS data and in both Visits D and E, showing the stability of the exosphere over many planetary orbits. No transit absorption was detected in the \ion{O}{i} line or in any of the other lines measured, with a global FUV transit depth of 3.3$\pm$3.8\%, preventing us from assessing whether hydrodynamical expansion sustains the giant exosphere of GJ\,3470b.

We used the synthetic XUV spectra to estimate the maximum mass loss and photoionization of GJ\,3470b upper atmosphere during quiescent and flaring epochs. Mass loss remained stable overall with an average maximum value of about 15$\times$10$^{10}$\,g\,s$^{-1}$ (considering the EMD based on the combined GJ\,3470 and AD~Leo data), and could have increased by up to 120\% and 320\% following Visit E and F flares, respectively. Theoretical studies are, however, required to assess whether the atmospheric outflow of warm Neptunes can be substantially affected by flares, especially from M dwarfs. The photoionization lifetime of neutral hydrogen atoms in GJ\,3470b exosphere decreased slightly over time following the increase in UV emission but remained close to about 40\,min. Flares had a direct impact on the exospheric structure, decreasing the neutral hydrogen lifetime down to 10\,min for the strongest episode, and thus changing the ionization fraction of the exosphere and possibly its interactions with the stellar magnetosphere. Future UV missions monitoring the GJ\,3470 system continuously over $\sim$2\,days could allow measuring the change in exospheric structure induced by a stellar flare.

%%%%%%%%%%%%%%%%%%%%%%%%%%%%%%%%%%%%%%%%%%%%%%%%%%%%%%%%%%%%%%%%%%%%%%%%%%%%%%%%%%%%%%%%%%%%%%%%%%%%%%%%%%%%%%%%%%%%%%%%%%

\begin{acknowledgements}
We thank the referee for their appreciative and careful review. This work is based on observations made with the NASA/ESA HST (PanCET program, GO 14767), obtained at the Space Telescope Science Institute (STScI) operated by AURA, Inc. This work has been carried out in the frame of the National Centre for Competence in Research “PlanetS” supported by the Swiss National Science Foundation (SNSF). This project has received funding from the European Research Council (ERC) under the European Union's Horizon 2020 research and innovation programme (project          {\sc Four Aces} grant agreement No 724427; project {\sc Spice Dune}; grant agreement No 947634). A.L. acknowledges financial support from the Centre National  d'\'Etudes Spatiales (CNES). G.W.H. acknowledges long-term support from NASA, NSF, Tennessee State University, and the State of Tennessee through its Centers of Excellence program.
\end{acknowledgements}

\bibliographystyle{aa} % style aa.bst
\bibliography{biblio} % your references Yourfile.bib

\begin{appendix}

\section{COS Lyman-$\alpha$ spectra of GJ\,3470}
\label{apn:ly_COS}

\begin{figure*}
\begin{minipage}[h!]{\textwidth}
\includegraphics[trim=3cm 1cm 3cm 3cm,clip=true,width=\columnwidth]{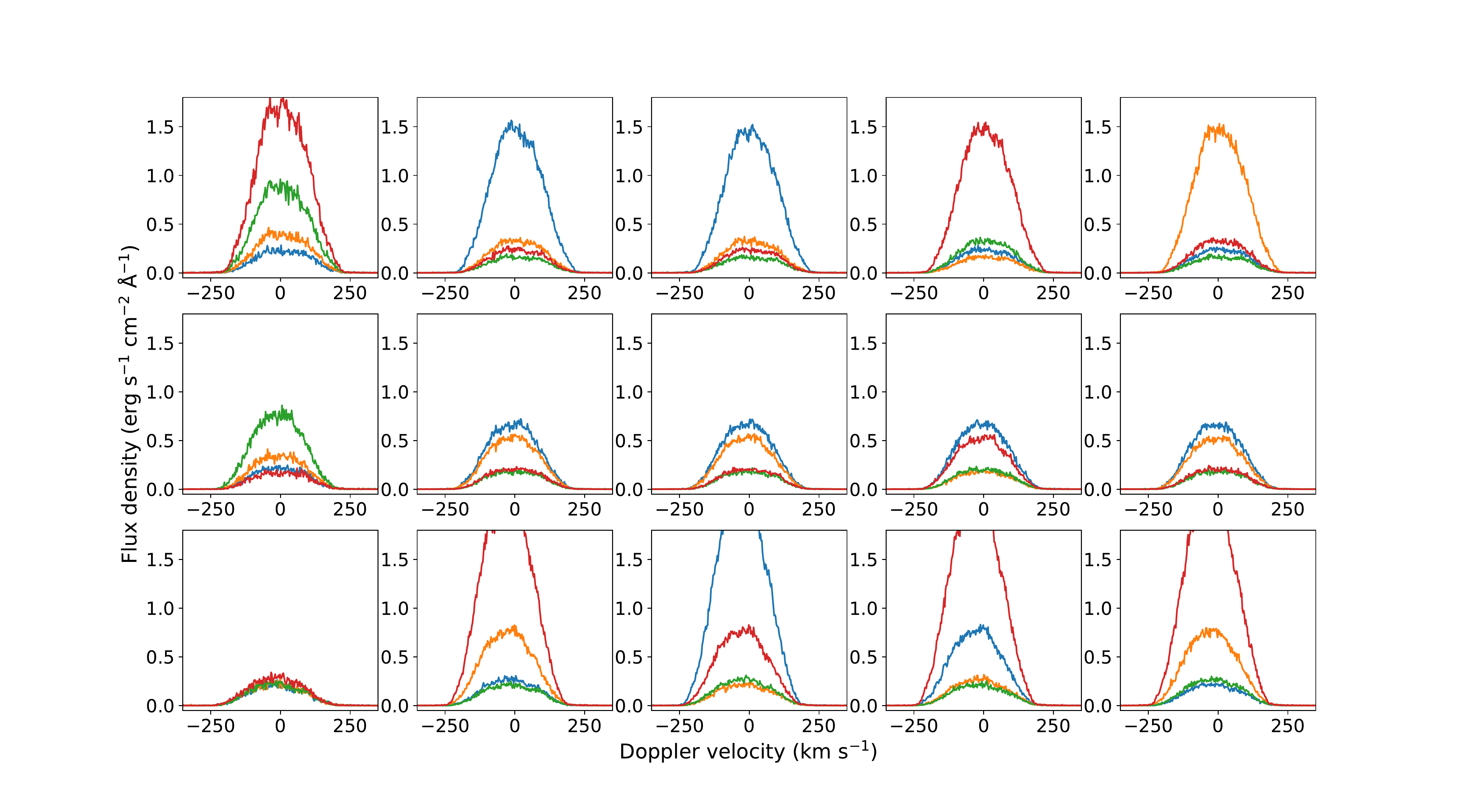}
\centering
\end{minipage}
\caption[]{Spectral profiles of GJ\,3470's Lyman-$\alpha$ line as observed with COS, still contaminated by geocoronal emission. From top to bottom, the rows correspond to Visits D, E, and F. Successive orbits in each visit are plotted from left to right. Each HST orbit is cut into four sub-exposures, plotted in blue, orange, green, and red, with increasing phase. Spectra are shown in the COS rest frame.}
\label{fig:Grid_Ly_airglow}
\end{figure*}

\section{Flare modeling}

\begin{figure*}
\begin{minipage}[h!]{\textwidth}
\includegraphics[trim=0cm 0cm 0cm 0cm,clip=true,width=0.7\columnwidth]{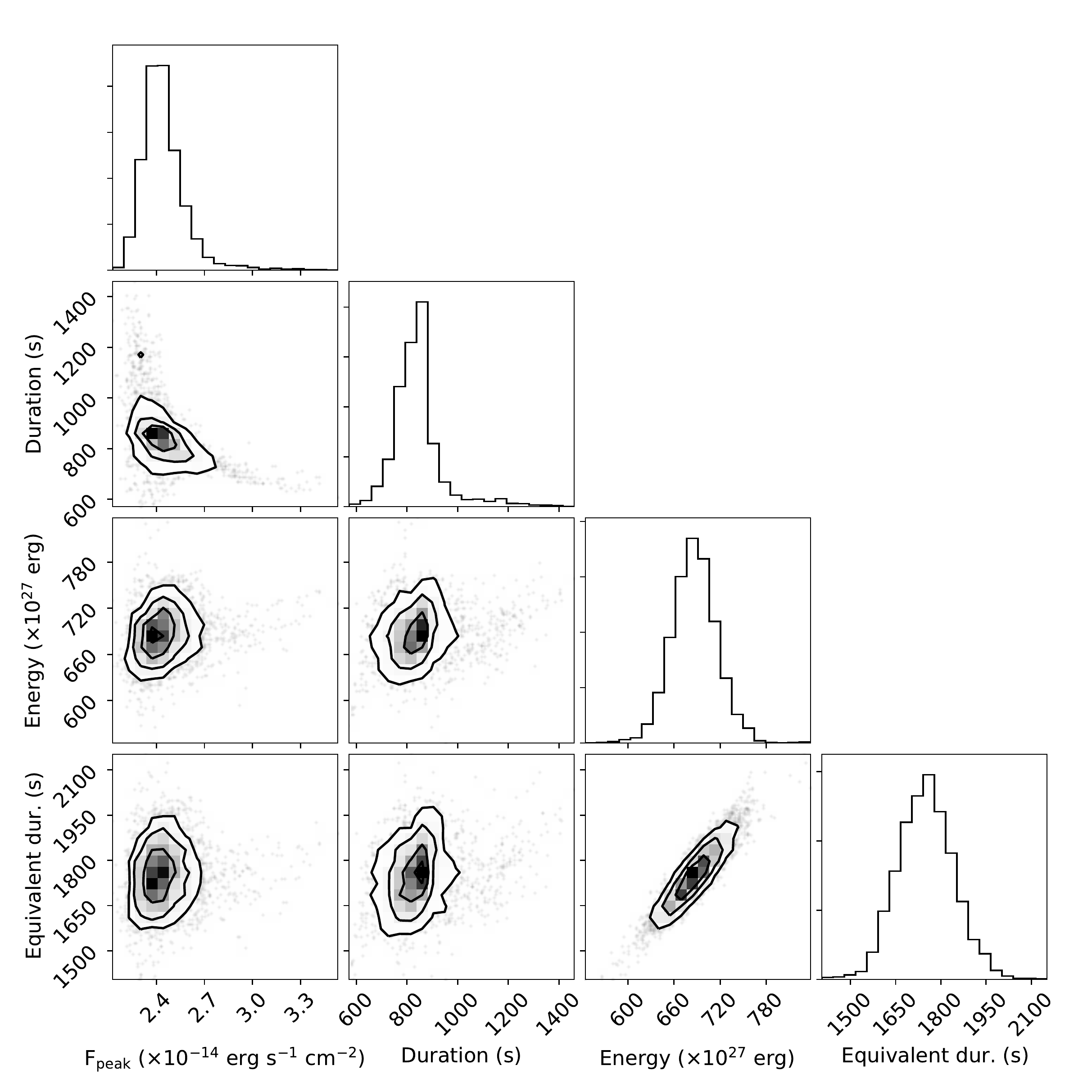}
\centering
\end{minipage}
\caption[]{Marginalized posterior distribution of the metrics for Visit F flare in the broadband FUV$_{130}$ bandpass.}
\label{fig:corner_flare_F130}
\end{figure*}

\section{Quiescent and flaring spectra of GJ\,3470}
\label{apn:GJ3470spec_grid}

\begin{figure*}
\begin{minipage}[h!]{\textwidth}
\includegraphics[trim=0cm 0cm 0cm 0cm,clip=true,width=0.8\columnwidth]{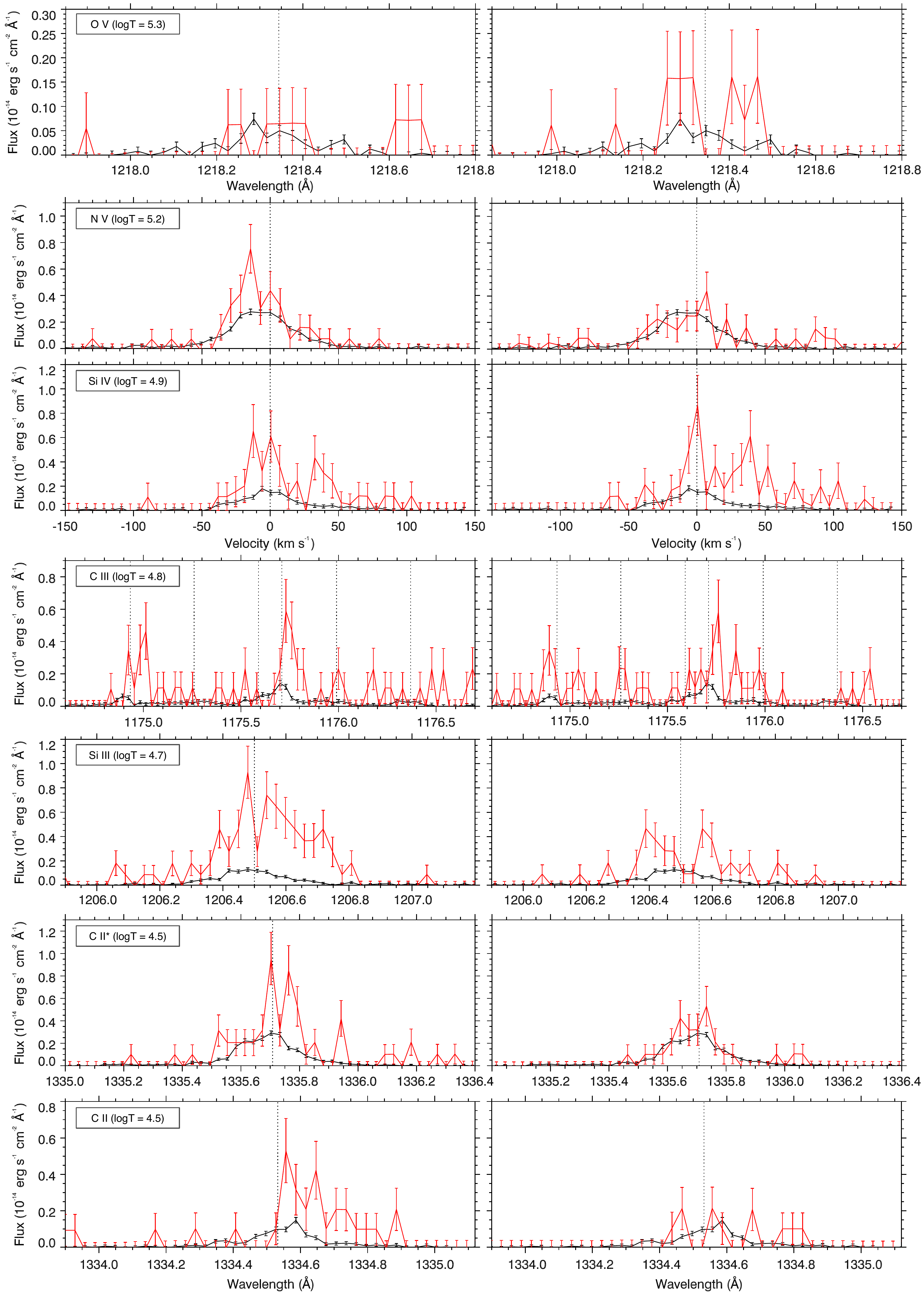}
\centering
\end{minipage}
\caption[]{Spectral profiles of GJ\,3470 FUV lines in Visit E. Each row corresponds to a stellar line, ordered from top to bottom by decreasing formation temperature. Black profiles show quiescent stellar lines. Red profiles correspond to the flaring exposures in the second (left column) and third (right column) HST orbits of the visit. Lines are plotted as a function of wavelength in the expected stellar rest frame. The \ion{N}{v} and \ion{Si}{iv} doublets have been coadded and are plotted in velocity space. Vertical dashed lines correspond to the expected rest position of the stellar lines. }
\label{fig:Flare_grid_spec_VE}
\end{figure*}

\begin{figure*}
\begin{minipage}[h!]{\textwidth}
\includegraphics[trim=0cm 0cm 0cm 0cm,clip=true,width=0.8\columnwidth]{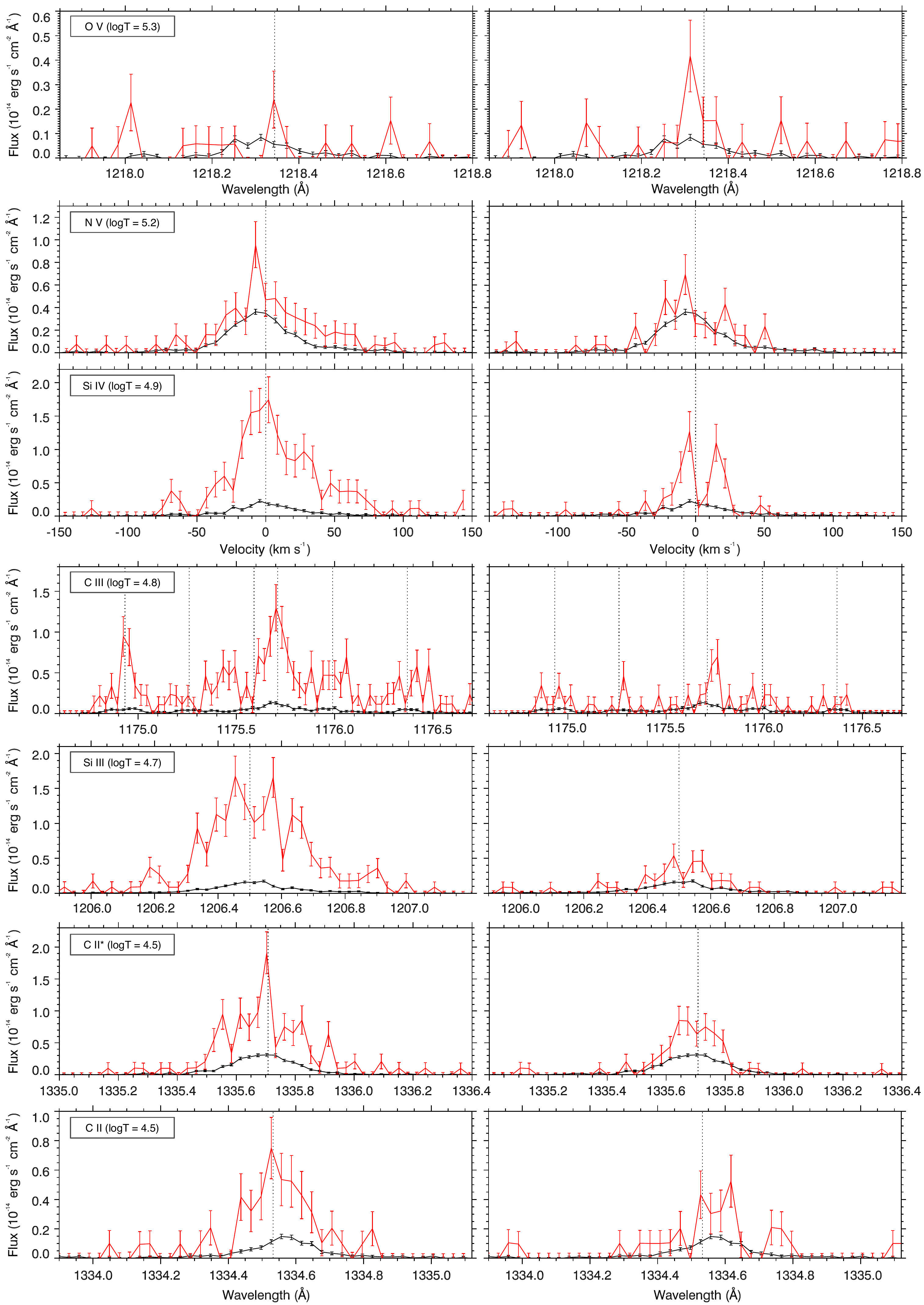}
\centering
\end{minipage}
\caption[]{Same as in Fig.~\ref{fig:Flare_grid_spec_VE}, but for Visit F. The left column shows the peak phase of the flare, while the right column shows its decay phase.}
\label{fig:Flare_grid_spec_VF}
\end{figure*}

\begin{figure*}
\begin{minipage}[h!]{\textwidth}
\includegraphics[trim=0cm 0cm 0cm 0cm,clip=true,width=\columnwidth]{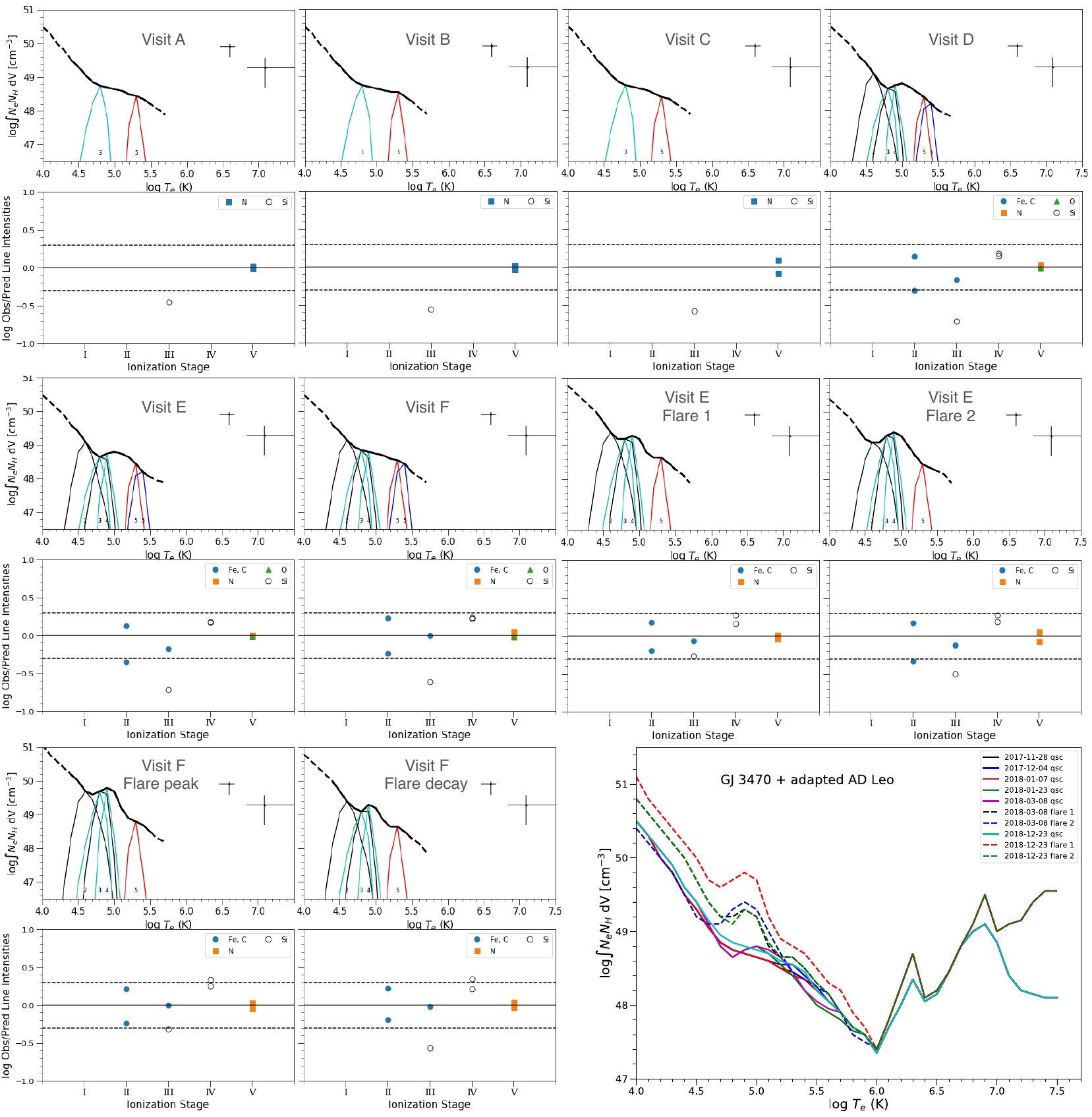}
\centering
\end{minipage}
\caption[]{EMDs of GJ\,3470 during the different stages. The coronal model is based on UV lines ($\log T<6.0$) and the X-ray XMM-Newton/EPIC spectral fit ($\log T>6.0$) from \citet{Bourrier2018_GJ3470b}. {\em Upper subpanels}: EMD indicated by thick black line based on UV lines and two discrete temperatures with error bars based on the X-ray spectral fit. Thin colored lines represent the emissivities of the different lines, weighted by the EMD, with ionization stages displayed in small numbers below the lines. {\em Lower subpanels}: Line flux ratios between the observed line fluxes and those predicted by the combination of the EMD and the atomic data, following \citet{Sanz-Forcada2011} and references therein. The plot in the lower right corner shows all EMDs together for the coronal model based on the combined GJ\,3470 and AD~Leo X-ray data, as explained in the text. We note that the AD~Leo data only provide a flaring and a quiescent stage.}
\label{fig:emdmixcomp}
\end{figure*}

%------------------------------------------------- begin table
%  HST/COS Line fluxes in all intervals
%\begin{landscape}
\begin{table*}
\caption[]{{\it HST} line fluxes of GJ 3470$^a$.}\label{tab:linefluxes} 
\tabcolsep 2 pt
\centering
%\begin{scriptsize}
\begin{tabular}{ccccccccccccccccc}
\hline \hline
%--------------------------------------------------------------
Ion & $\lambda_{\rm model}$ & \multicolumn{3}{c}{2017-11-28 qsc} & \multicolumn{3}{c}{2017-12-04 qsc} & \multicolumn{3}{c}{2018-01-07 qsc} & \multicolumn{3}{c}{2018-01-23 qsc} & \multicolumn{3}{c}{2018-03-08 qsc} \\
 & (\AA) & $F_{\rm obs}$ & $S/N$ & Ratio & $F_{\rm obs}$ & $S/N$ & Ratio & $F_{\rm obs}$ & $S/N$ & Ratio & $F_{\rm obs}$ & $S/N$ & Ratio & $F_{\rm obs}$ & $S/N$ & Ratio  \\
\hline
%--------------------------------------------------------------
\ion{Si}{iii} & 1206.5019 & 5.49e-16 & 4.9 & -0.45 & 4.32e-16 & 4.5 & -0.56 & 4.13e-16 & 4.3 & -0.58 & 3.11e-16 & 22.9 & -0.71 & 3.08e-16 & 22.8 & -0.72 \\
\ion{O}{v} & 1218.344 & \dots & \dots & \dots & \dots & \dots & \dots & \dots & \dots & \dots & 1e-16 & 9.1 & -0.01 & 1.11e-16 & 8.7 & -0.01 \\
\ion{N}{v} & 1238.8218 & 4.87e-16 & 7.4 & 0.02 & 5.16e-16 & 7.3 & -0.03 & 3.69e-16 & 6.6 & -0.08 & 4.48e-16 & 39.3 & 0.01 & 5.11e-16 & 33.8 & 0.00 \\
\ion{N}{v} & 1242.8042 & 2.26e-16 & 4.8 & -0.01 & 2.89e-16 & 5.9 & 0.03 & 2.76e-16 & 5.8 & 0.10 & 2.33e-16 & 25.5 & 0.03 & 2.57e-16 & 17.6 & 0.01 \\
\ion{C}{ii} & 1334.535 & \dots & \dots & \dots & \dots & \dots & \dots & \dots & \dots & \dots & 5.43e-16 & 17.6 & -0.31 & 5.02e-16 & 17.9 & -0.35 \\
\ion{C}{ii} & 1335.71 & \dots & \dots & \dots & \dots & \dots & \dots & \dots & \dots & \dots & 6.95e-16 & 41.6 & 0.14 & 6.72e-16 & 35.2 & 0.13 \\
\ion{Si}{iv} & 1393.7552 & \dots & \dots & \dots & \dots & \dots & \dots & \dots & \dots & \dots & 2.53e-16 & 20.7 & 0.15 & 2.75e-16 & 20.2 & 0.18 \\
\ion{Si}{iv} & 1402.7704 & \dots & \dots & \dots & \dots & \dots & \dots & \dots & \dots & \dots & 1.36e-16 & 11.4 & 0.18 & 1.36e-16 & 10.4 & 0.17 \\
\ion{C}{iii} & 1176.0 & \dots & \dots & \dots & \dots & \dots & \dots & \dots & \dots & \dots & 2.02e-15 & 10.7 & 0.42 & 5.19e-16 & 23.5 & -0.17 \\
\hline
\hline
%--------------------------------------------------------------
%--------------------------------------------------------------
 & & \multicolumn{3}{c}{2018-03-08 flare 1} & \multicolumn{3}{c}{2018-03-08 flare 2} & \multicolumn{3}{c}{2018-12-23 qsc} & \multicolumn{3}{c}{2018-12-23 flare peak} & \multicolumn{3}{c}{2018-12-23 flare decay} \\
\hline
%--------------------------------------------------------------
\ion{Si}{iii} & 1206.5019 & 2.3e-15 & 17.3 & -0.27 & 1.15e-15 & 11.2 & -0.50 & 4.76e-16 & 28.9 & -0.61 & 5.26e-15 & 23.9 & -0.32 & 1.1e-15 & 9.1 & -0.56 \\
\ion{N}{v} & 1238.8218 & 7.2e-16 & 8.1 & 0.01 & 5.08e-16 & 6.5 & -0.07 & 5.56e-16 & 54.0 & -0.02 & 1.23e-15 & 17.4 & 0.03 & 7.62e-16 & 12.9 & 0.04 \\
\ion{N}{v} & 1242.8042 & 3.21e-16 & 4.7 & -0.04 & 3.35e-16 & 5.0 & 0.05 & 3.2e-16 & 31.4 & 0.04 & 5.11e-16 & 10.3 & -0.05 & 3.29e-16 & 7.2 & -0.03 \\
\ion{C}{ii} & 1334.535 & 1.62e-15 & 7.3 & -0.19 & 6.02e-16 & 3.6 & -0.34 & 6.93e-16 & 25.8 & -0.24 & 2.94e-15 & 13.5 & -0.24 & 1.59e-15 & 9.1 & -0.19 \\
\ion{C}{ii} & 1335.71 & 1.72e-15 & 89.8 & 0.18 & 8.53e-16 & 7.2 & 0.17 & 9.2e-16 & 50.0 & 0.23 & 3.64e-15 & 22.2 & 0.21 & 1.84e-15 & 14.7 & 0.22 \\
\ion{Si}{iv} & 1393.7552 & 8.01e-16 & 8.2 & 0.17 & 1.06e-15 & 10.3 & 0.19 & 3.55e-16 & 31.2 & 0.23 & 3.01e-15 & 28.2 & 0.25 & 1.18e-15 & 15.2 & 0.35 \\
\ion{Si}{iv} & 1402.7704 & 5.15e-16 & 6.0 & 0.28 & 6.41e-16 & 6.7 & 0.28 & 1.73e-16 & 15.8 & 0.22 & 1.81e-15 & 20.4 & 0.33 & 4.27e-16 & 6.9 & 0.21 \\
\ion{C}{iii} & 1176.0 & 1.91e-15 & 10.9 & -0.07 & 1.8e-15 & 10.5 & -0.12 & 9.54e-16 & 36.1 & -0.00 & 6.42e-15 & 22.1 & 0.00 & 2.02e-15 & 10.7 & -0.02 \\
%--------------------------------------------------------------
\hline

\end{tabular}

\begin{tablenotes}[para,flushleft]
$^a$ Line fluxes (in erg cm$^{-2}$ s$^{-1}$) measured in {\it HST}/STIS and COS GJ 3470 spectra, and corrected for the ISM absorption when relevant. log $T_{\rm max}$ (K) indicates the maximum temperature of formation of the line (unweighted by the EMD). ``Ratio'' is the log($F_{\mathrm {obs}}$/$F_{\mathrm {pred}}$) of the line.
 
\end{tablenotes}   

%\end{scriptsize}
%\end{centering}
%\renewcommand{\arraystretch}{1.}
\end{table*}
%\end{landscape}
%--------------------------------------------- end table

%------------------------------------------------- begin table
%  Emission Measure Distribution
\begin{table*}
\caption{Emission measure distribution of GJ 3470 in the different intervals.}\label{tab:emd}
%\tabcolsep 3.pt
\begin{center}
\begin{small}
\begin{tabular}{ccccccccccc}
\hline \hline
{$\log T$} & \multicolumn{10}{c}{$\log EM$ (cm$^{-3}$)} \\
(K) & A & B & C & D & E & E$_{\rm f1}$ & E$_{\rm f2}$ & F & F$_{\rm fp}$ & F$_{\rm fd}$ \\
\hline
%----------------
4.0 & 50.50: & 50.50: & 50.50: & 50.50: & 50.50: & 50.80: & 50.40: & 50.50: & 51.10: & 50.80: \\
4.1 & 50.30: & 50.30: & 50.30: & 50.30: & 50.30: & 50.60: & 50.20: & 50.30: & 50.80: & 50.60: \\
4.2 & 50.00: & 50.00: & 50.00: & 50.10: & 50.10: & 50.40: & 50.00: & 50.10: & 50.60: & 50.40: \\
4.3 & 49.80 & 49.80 & 49.80 & 49.90 & 49.90 & 50.20 & 49.80 & 49.90 & 50.40 & 50.20 \\
4.4 & 49.50 & 49.50 & 49.50 & 49.60 & 49.60 & 50.00 & 49.50 & 49.60 & 50.20 & 50.00 \\
4.5 & 49.30 & 49.30 & 49.30 & 49.40 & 49.40 & 49.70 & 49.20 & 49.40 & 50.00 & 49.70 \\
4.6 & 49.05 & 49.05 & 49.05 & 49.10 & 49.10 & 49.40 & 49.10 & 49.15 & 49.70 & 49.40 \\
4.7 & 48.85 & 48.85 & 48.85 & 48.80 & 48.80 & 49.20 & 49.10 & 48.95 & 49.60 & 49.20 \\
4.8 & 48.75 & 48.75 & 48.75 & 48.65 & 48.65 & 49.20 & 49.30 & 48.85 & 49.70 & 49.10 \\
4.9 & 48.70 & 48.70 & 48.70 & 48.75 & 48.75 & 49.30 & 49.40 & 48.80 & 49.80 & 49.30 \\
5.0 & 48.65 & 48.65 & 48.65 & 48.80 & 48.80 & 49.20 & 49.30 & 48.75 & 49.70 & 49.20 \\
5.1 & 48.60 & 48.60 & 48.60 & 48.70 & 48.75 & 48.80 & 49.00 & 48.70 & 49.20 & 48.85 \\
5.2 & 48.50 & 48.55 & 48.50 & 48.55 & 48.65 & 48.65 & 48.70 & 48.60 & 48.90 & 48.65 \\
5.3 & 48.45 & 48.55 & 48.40 & 48.40 & 48.45 & 48.65 & 48.45 & 48.55 & 48.80 & 48.65 \\
5.4 & 48.35 & 48.40 & 48.35 & 48.20 & 48.20 & 48.50 & 48.35 & 48.45 & 48.70 & 48.50 \\
5.5 & 48.20 & 48.25 & 48.20 & 48.00 & 48.05 & 48.30 & 48.25 & 48.20 & 48.50 & 48.30 \\
5.6 & 48.05: & 48.05: & 48.05: & 47.90: & 47.95: & 48.15: & 48.15: & 48.05: & 48.30: & 48.15: \\
5.7 & 47.90: & 47.90: & 47.90: & 47.80: & 47.90: & 47.90: & 47.90: & 47.90: & 48.20: & 47.90: \\
5.8 & 45.90: & 45.90: & 45.90: & 45.90: & 45.90: & 45.90: & 45.90: & 45.90: & 46.10: & 45.90: \\
5.9 & 46.00: & 46.00: & 46.00: & 46.00: & 46.00: & 46.00: & 46.00: & 46.00: & 46.20: & 46.00: \\
%---------------
\hline
\end{tabular}
\end{small}
\end{center}

\begin{tablenotes}[para,flushleft]
Notes: Emission measure ($EM= \int N_{\rm e} N_{\rm H} {\rm d}V$), where $N_{\rm e}$ and $N_{\rm H}$ are electron and hydrogen densities, in cm$^{-3}$.
\end{tablenotes}    
    
\end{table*}

\begin{figure*}
\begin{minipage}[h!]{\textwidth}
\includegraphics[trim=0cm 0cm 0cm 0cm,clip=true,width=\columnwidth]{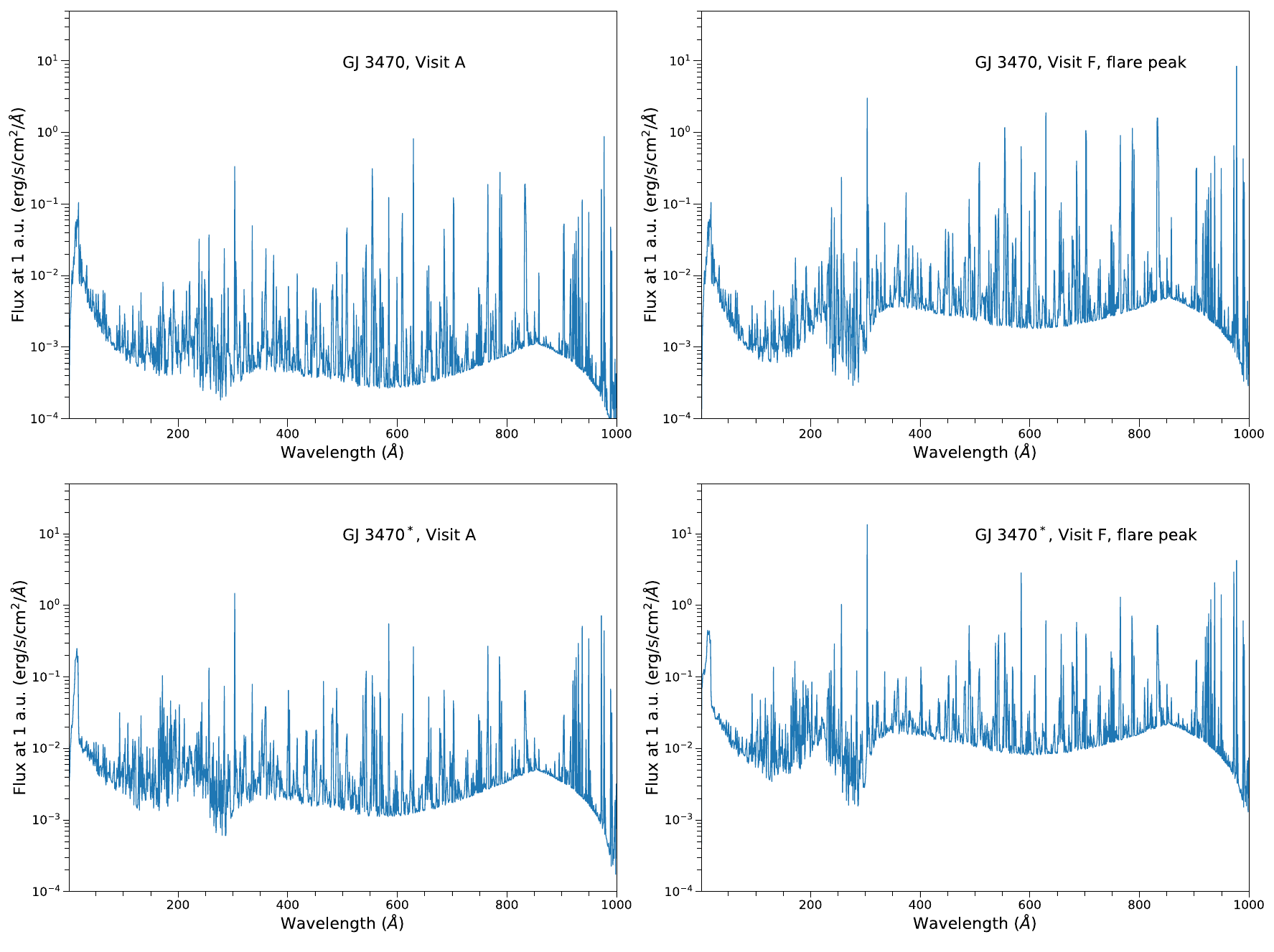}
\centering
\end{minipage}
\caption[]{Spectral energy distribution as modeled using the EMD in Fig.~\ref{fig:emdmixcomp}, shown here for the quiescent state in Visit A (left panel) and the flaring peak phase in Visit F (right panels). Upper panels use the coronal data from the XMM-Newton observation of GJ 3470 alone. Lower panels use the coronal EMD adapted from AD Leo quiescent and flaring stages. The strongest line at high energies corresponds to \ion{He}{ii}~304\,\AA.}
\label{fig:EUV_spectra}
\end{figure*}

\end{appendix}

\end{document}